\documentclass[aps,prstab,twocolumn,groupedaddress]{revtex4}
\usepackage{amsmath}
\usepackage{subfigure}
\usepackage{epsfig}

\begin{document}

%\preprint{}

\title{Influence of a magnetic guide field on wakefield acceleration}


\author{M. Drouin}
\email{mathieu.drouin@cea.fr}
\affiliation{CEA, DAM, DIF, F-91297 Arpajon, France}

\author{Q. Harry}

\author{A. Bourdier}
\email{alain.bourdier@cea.fr}
\affiliation{CEA, DAM, DIF, F-91297 Arpajon, France}

%\thanks{}
\affiliation{}


\date{August 22, 2011}

\begin{abstract}
Enhancement of the trapping and optimization of the beam quality are two key issues of Laser Wake Field Acceleration (LWFA).
The influence of stochastic acceleration on the trapping of electrons is compared to the one of cold injection.
It is shown that when considering a high intensity wave perturbed by a low intensity
counter-propagating wave, in the non-linear
blowout regime, the influence of the colliding pulses polarizations (either parallel linear or positive circular)
on the beam quality seems weak when the electron density
is below $\sim 10^{-3}$ critical density.
The effect of a homogenous constant magnetic field $B_0$, parallel to the direction of propagation of the pump pulse, is studied
in the blowout regime. Transverse currents are generated at the rim of the bubble, which results in the amplification of the
$B_0$ field at the rear of the bubble.
Without $B_0$ field the beam periodically explodes and re-confines, this phenomenon is suppressed when $B_0$ reaches some
threshold, which is a function of the laser pulses parameters (intensity, waist, duration).
Therefore the dynamics of the beam is modified, its maximum energy is slightly boosted and above all transverse emittance
reduced. Moreover the low energy tail, observed in the non magnetized case, can be completely suppressed leading to
very sharp mono-energetic beam when $B_0$ is applied.
If the available $B_0$ field is limited then one has to fine-tune the spatio-temporal shape and intensity of the colliding
pulse in order to get an acute control on the beam quality.
\end{abstract}

\pacs{52.38.Kd}
%\keywords{}

\maketitle

%\section{}
\section{\label{sec-intro}Introduction}
In laser-wakefield acceleration (LWFA) \cite{tajima1979, esarey_prl1997, faure2004, mangles2004}, a laser creates a plasma wave wakefield with a
phase velocity close to the speed of light ($c$). The acceleration gradients in these wakefields
can easily exceed 100 GeV/m, hence a cm-long plasma based accelerator can produce GeV-energy
electron beams. A particle injected in such a wave gains energy from the longitudinal component
of the electric field, as long as the pump pulse is not depleted and the dephasing length
is not reached. These wakefields have ideal properties for accelerating electrons. The
transverse focusing field increases linearly with the radial distance and the accelerating
longitudinal field is independent of the radial coordinate \cite{rosenz_1991,lu_prl2006}.
LWFA can be split into three different options .
The first corresponds to a plasma density
$n_e \approx 10^{19} \textrm{cm}^{-3}$, a pulse length ($c\tau$) matching half of a
plasma period and a spot size ($w_0$) roughly equals to the bubble radius,
$w_0 \approx c\tau \approx \sqrt{a_0}$, where $a_0$ is the normalized vector potential of the laser.
This is the idea of the bubble regime \cite{pukhov_2002,gordienko_2005}.
For these conditions, a hundred-joule class laser would have an intensity of the order
$\sim 10^{21} \textrm{W cm}^{-2}$.
In this regime, the electrons are continuously injected, this results in tremendous beam
loading and the loaded wake is noisy.
In this paper we explore different techniques to improve the beam quality of LWFA when electrons are
injected in the wake with a colliding pulse. Hence, the bubble regime is not appropriate. We rather
select moderate laser intensity $I \le 10^{19} \textrm{W.cm}^{-2}$ and plasma density
$n_e \le 10^{18} \textrm{cm}^{-3}$ according to the
guidelines proposed by Lu et al.\cite{lu_prstab2007} to achieve a more controlled and stable blowout of the
electrons.
Self-injection of electrons can occur when the pump pulse intensity is high but the
accelerating structure is almost the same. In this frame beam loading effects clamp further
injection leading to beams with a smaller energy spread. In order to limit the computational
requirements of our PIC simulations the propagation of the pump pulse will not exceed 1 cm.
We tend to avoid self-injection into the wake by adjusting the pump pulse intensity and
the electronic density.
Many different combinations of polarizations can be
chosen for both waves, each of these possibilities results in particular force acting
on the plasma electrons, when the two waves collide \cite{davoine_2008}.
The first point of this article is to summarize the dependance to pump and colliding pulse intensities,
and to plasma density of this force.
The relative influence of stochastic heating and beat wave force on the injection mechanism,
and later on the beam quality will be discussed.
After the choice of polarization in the blowout regime is clarified, we focus on the study
of wakefield acceleration in the presence of an external, homogenous, magnetic field and
study its influence through simulations. The mechanisms leading to the enhancement of the
beam quality will be examined.
In a third part, fine-tuning of the counter-propagating low intensity pulse will be considered
in order to limit the intensity of the external field.
This situation will be illustrated in the case when the intensity of the pump pulse is raised
to $a_0=10$. Then we will conclude.
\section{Sensitivity of beam injection to wave intensity, plasma density and polarisation of lasers}
\subsection{Basic principles\label{subsec:principles}}

The wakefield propagates in the plasma at the group velocity of the laser $\beta_g$ defined by :
$\beta_g = v_g/c =\left( 1 - \frac{\omega_p^2}{\omega_0^2} \right)^{1/2}$,
where $c$ is the speed of light, $\omega_p$ and $\omega_0$ respectively denote the plasma and
laser frequencies. We use the quasi-static approximation and assume that the potential $\phi$ created
by the pump pulse only depends on $\xi=x-\beta_g t$, where $x$ and $t$ denote the space and time
coordinates normalized by $c/\omega_0$ and $\omega_0^{-1}$ respectively.
Then, the hamiltonian of an electron in the wakefield potential $\phi$, created by the pump pulse, reads :
\begin{equation}
H(\xi, p_x)=\sqrt{\gamma_{\perp}^2 + p_x^2} - \beta_g p_x -\phi(\xi),
\label{eq_hamil01}
\end{equation}
where $\gamma_{\perp}^2=1+\mathbf{p}_{\perp}^2$.
The normalized transverse momentum is defined by $\mathbf{p}_{\perp}=\frac{\mathbf{u}_{\perp}}{m_e c}$, where $\mathbf{u}_{\perp}$
is the transverse momentum and $m_e$ denotes the electron mass.
The hamiltonian of a particle is an invariant $H(\xi,p_x)=H_0$, using (\ref{eq_hamil01}) we deduce
two solutions for the longitudinal momentum :
\begin{equation}
\label{eq_hamil02}
p_{x}^{\pm}(\xi)=\beta_g \gamma_g^2(H_0+\phi) \pm \gamma_g \left(\gamma_g^2(H_0 + \phi)^2 - \gamma_{\perp}^2\right)^{1/2},
\end{equation}
where $\gamma_g^2=1+\beta_g^2$.
The separatrix between trapped and untrapped orbits is given by a critical value,
$H_s=\frac{\gamma_{\perp}}{\gamma_g}-\phi_{\textrm{min}}$, of the Hamiltonian. Replacing $H_0$ by $H_s$
in (\ref{eq_hamil02}) and retaining $p_x^{\pm}$ we get the two branches of $p_x^{\textrm{sep}}(\xi)$.
%=\beta_g \gamma_g^2(H_s+\phi)-\gamma_g\left( \gamma_g^2(H_s+\phi)^2-\gamma_{\perp}^2\right)^{1/2}

Let us now comment the dynamics of an electron in vacuum, in the presence of two counter-propagating
laser pulses.
The hamiltonian reads $H(u_x,x)=\left(1+u_x^2+u_{\perp}^2\right)^{1/2}=\gamma$, the longitudinal force
acting on the electron is given by
\begin{equation}
\label{eq_hamil03}
\frac{du_x}{dt}=-\frac{\partial H}{\partial x}= - \frac{1}{2\gamma}\frac{\partial u_{\perp}^2}{\partial x}.
\end{equation}
Taking only into account the influence of the lasers, denoted by their potential vectors $\textbf{A}_0$
and $\textbf{A}_1$, we have $\textbf{u}_{\perp}=\textbf{A}_0+\textbf{A}_1$.
We will consider parallel linear polarization (P linear polarization) and positive circular polarization, that is
$\textbf{u}_{\perp}=a_0 \textrm{cos}(\omega_0 t - k_0 x)\textbf{e}_y + a_1 \textrm{cos}(\omega_0 t + k_0 x)\textbf{e}_y$
and $\textbf{u}_{\perp}=\left[ \frac{a_0}{\sqrt{2}}\textrm{cos}(\omega_0 t - k_0 x) + \frac{a_1}{\sqrt{2}}\textrm{cos}(\omega_0 t + k_0 x)\right] \textbf{e}_y +
\left[ \frac{a_0}{\sqrt{2}}\textrm{sin}(\omega_0 t - k_0 x) + \frac{a_1}{\sqrt{2}}\textrm{sin}(\omega_0 t + k_0 x)\right] \textbf{e}_z$ respectively.
Substituting the above expressions in Eq. (\ref{eq_hamil03}) yields
\begin{align}
\label{eq_hamil04}
F_x & =\frac{-1}{2\gamma}\left[ a_0^2 k_0 \textrm{sin}(2(\omega_0 t - k_0 x)) \right. \notag \\
    & \left. - a_1^2 k_0 \textrm{sin}(2(\omega_0 t + k_0 x)) \right] + \frac{k_0 a_0 a_1}{\gamma}\textrm{sin}(2k_0 x),
\end{align}
for the P linear polarization case, and
\begin{equation}
\label{eq_hamil05}
F_x= \frac{k_0 a_0 a_1}{\gamma}\textrm{sin}(2k_0 x) \equiv F_{bw},
\end{equation}
for the positive circular polarization case.
When P linear or positive circular polarizations are used Eqs. \eqref{eq_hamil04} and \eqref{eq_hamil05} show the existence
of a force $F_{bw}$ spatially oscillating with a $\lambda_0/2$ period.
This force is not time-dependant, it is usually interpreted as a ponderomotive force associated with the beat-wave
\cite{fubiani_2004,kotaki_2004,faure_2006,davoine_2008}.
We should compare the beatwave force $F_{bw}$ to the longitudinal ponderomotive force, this latter scales as
$F_p = (2\gamma)^{-1}a_0^2/(c\tau)$ where $\tau$ is the pulse duration.
Taking the maximum value of $F_{bw}$ the ratio $F_{bw}/F_p$ becomes
\begin{equation}
F_{bw}/F_p = 2 a_1 \omega_0 \tau /a_0.
\end{equation}
When $F_{bw}/F_p>1$ electrons are trapped inside the $\lambda_0/2$-long beatwave buckets and cannot be wiped out
by the longitudinal ponderomotive force.
On the whole for both polarizations electrons will undergo the beatwave force, therefore when the separatrix is such that
$\min(u_x)<0$ electrons will be trapped in the wakefield as a bunch \cite{davoine_prl2009}.
However in the P linear polarization case other terms are added  to the force and the equations of the motion
are no longer integrable, electron trajectories become chaotic \cite{bourdier_phyD2005}.
This phenomenon known as stochastic heating can provide very large momenta to some electrons
\cite{sheng2002, sheng2004,  patin_2005a, patin_lpb2006, bourdier_lpb2007, bourdier_lpb2009, bourdier_ieee2010}.
When $\min(u_x)\ge 0$ the beatwave force is not efficient and electrons can hardly be trapped with positive circular
polarizations. In this case P linearly polarized colliding waves are necessary to give electrons the appropriate momentum in order to
fill the gap between trapped and untrapped orbits. In the following two subsections, the sensibility of the injection
process to polarization and intensity of the waves will be summarized.

\subsection{Low density: Strong dependance to polarization and wave intensity}
\begin{figure}[!htbp]
\begin{center}
  \subfigure[$a=1.5$, $a_1=0.4$]
  {\epsfig{figure=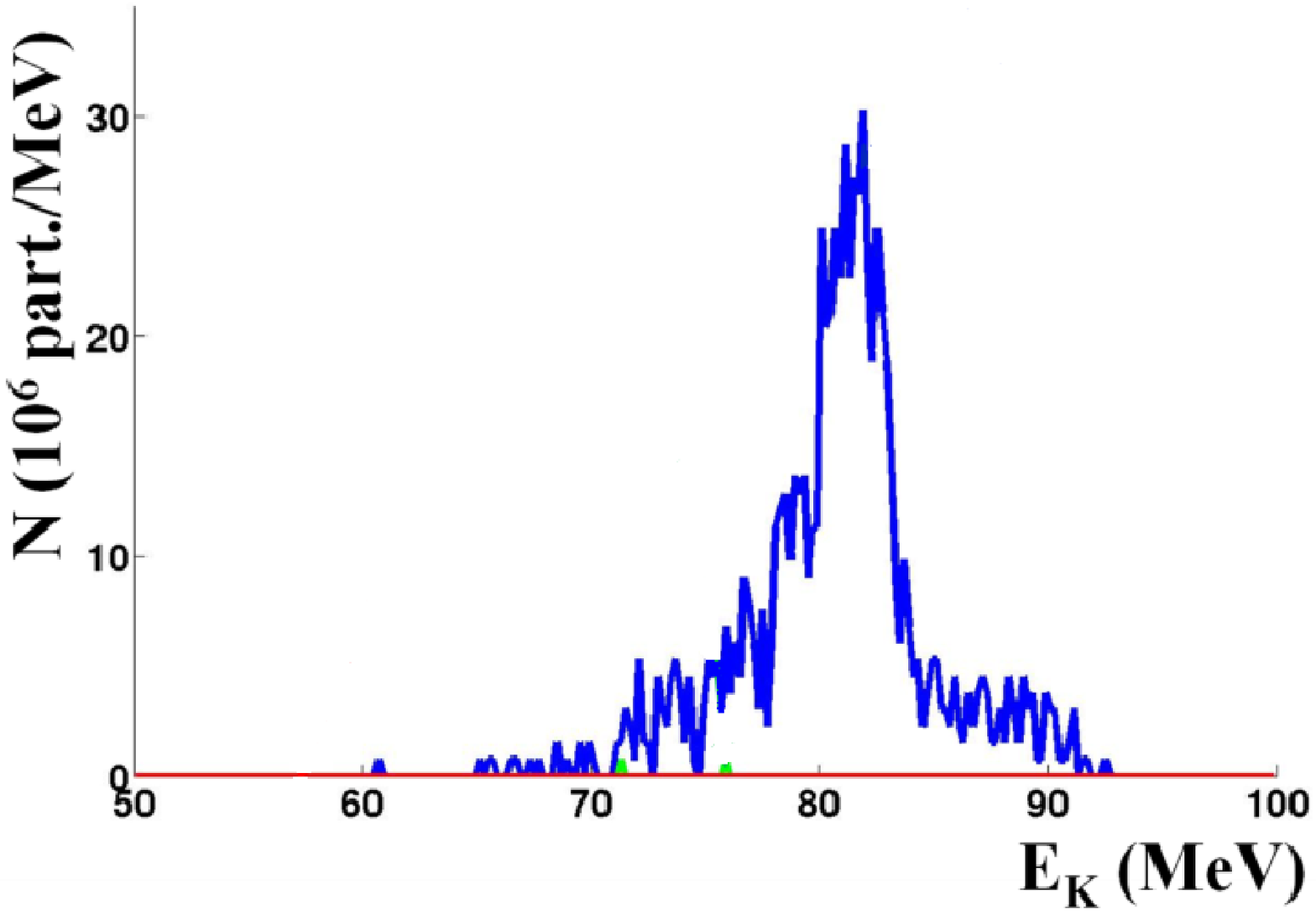,height=2.5cm}
  \label{fig_f_E_7000_weak_01}}
  \subfigure[$a=2$, $a_1=0.4$]
  {\epsfig{figure=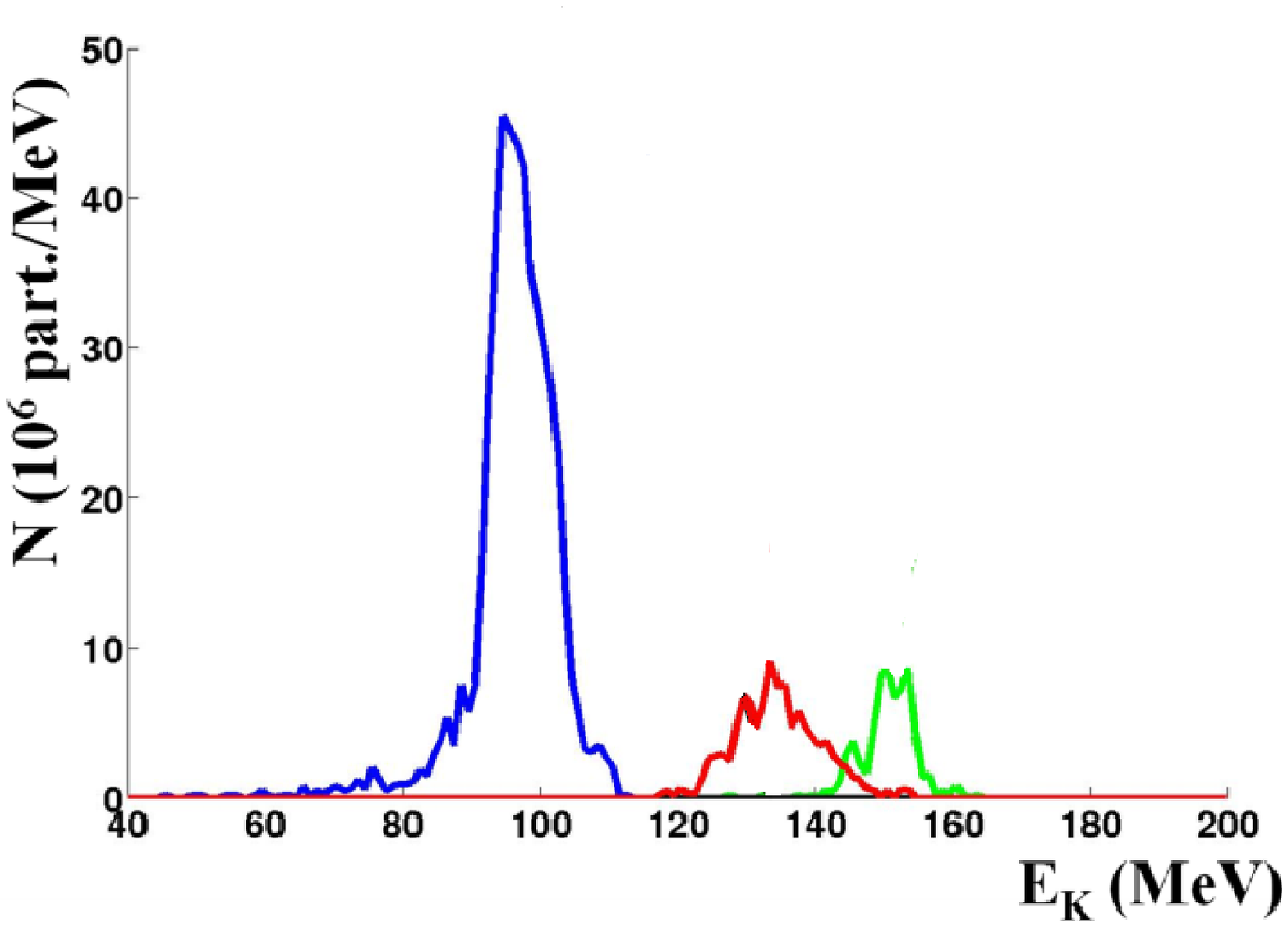,height=2.5cm}
  \label{fig_f_E_7000_weak_02}}
  \subfigure[$a=2$, $a_1=0.1$]
  {\epsfig{figure=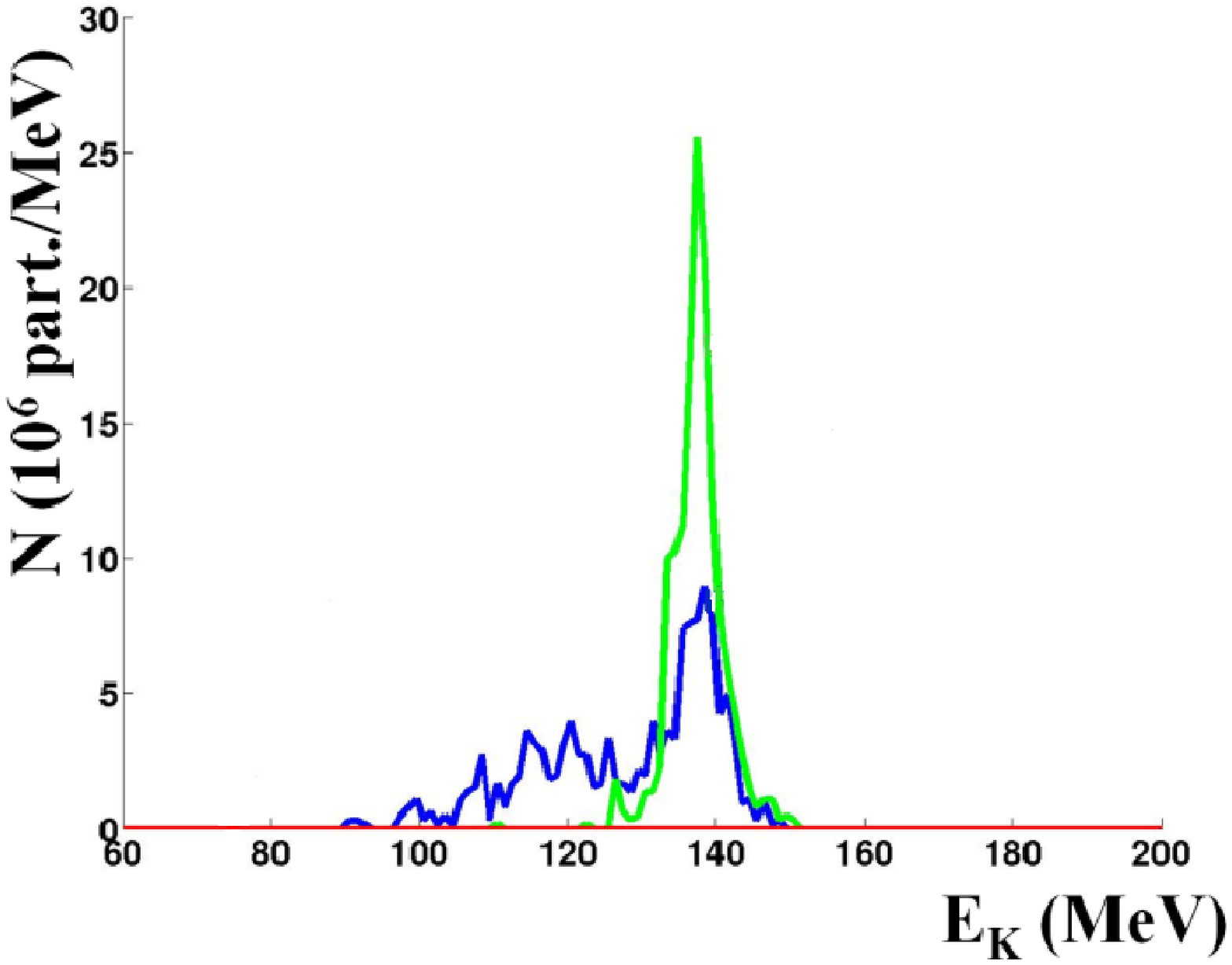,height=2.5cm}
  \label{fig_f_E_7000_weak_03}}
\end{center}
\caption{Electron energy distribution obtained long after the collision of the two waves
with $n_e = 4.3 \times 10^{-3} n_c$.
P linear polarizations (blue curve). Positive circular polarizations (green curve).
S linear polarizations (red curve) and negative circular polarizations
(black curve, overlaid with axis in this case).
\label{fig_f_E_7000_weak}}
\end{figure}

When the plasma density is about $\sim 10^{-3} n_c$ a weak variation of the pump pulse is sufficient to
modify the mechanisms allowing trapping of a bucket in the wakefield.
To illustrate this strong dependance to laser parameters we launched three sets of simulations
with the PIC code CALDER \cite{lefebvre_2003}.
The simulation setup consists in two 30 fs linearly polarized waves with wavelength $\lambda = 0.8 \mu$m
having their electric fields either linearly or circularly polarized (orthogonal linear polarization is
denoted by S).
The pump pulse, which creates the accelerating wakefield, is focused to an $18 \mu$m full width at half maximum (fwhm).
The peak normalized vector potential of the main pulse is associated
to the laser intensity $I$ by the formula $a = 0.853\times10^{-9} \lambda(\mu \textrm{m})[I(\textrm{W cm}^{-2})]^{1/2}$,
we considered $a=1.5$ and $a=2$.
The low intensity pulse is counter-propagating and is focused to a $31 \mu$m focal spot at a peak normalized
vector potential $a_1=0.1$ or $0.4$. The waves interact with a mm-size plasma with a density
$n_e=4.3\times10^{-3} n_c$.
%\begin{figure}[htbp]
%\includegraphics[width=0.25\textwidth]{qxqy_t6060_43e-4}
%\caption{Electron density long after the collision of the two waves. P linear polarizations.
%$a=1.5$, $a_1=0.4$ and $n_e = 4.3\times 10^{-3} n_c$.}
%\label{fig_weak_qxqy01}
%\end{figure}
The fluctuations between the two regimes are illustrated by Figs. \ref{fig_f_E_7000_weak}.
When $a=1.5$  and $a_1=0.4$ the $1D$ separatrix between trapped and opened orbits is higher than $p_x = 0$, in this case
the beatwave force cannot provide enough momentum to electrons to push them in the wakefield.
Nevertheless stochastic heating due to P linear polarizations is a way to bridge the gap and inject
a bunch in the wakefield (Fig. \ref{fig_f_E_7000_weak_01}).
When $a=2$ the $1D$ separatrix is lowered ($\min (p_x) < 0$), in this case the beatwave force is enough to trap
electrons in the wake therefore we can accelerate a beam using either P linear or positive circular polarizations.
Note that no trapping occurs with negative circular polarizations, which is consistent with the theory
introduced in subsection \ref{subsec:principles}. Indeed straightforward algebra gives $F_x=0$, then the ponderomotive force
hinders trapping of electrons.
The quality of the accelerated beam will depend on the force which dominates during collision of the pulses.
A relatively high intensity of the counter propagating laser ($a_1=0.4$) will foster stochastic heating
\cite{bourdier_lpb2009, bourdier_ieee2010} and a higher charge will be injected with P linear polarizations
compared to positive circular polarizations (Fig. \ref{fig_f_E_7000_weak_02}).
On the contrary, if we reduce the intensity of the counter propagating pulse ($a_1=0.1$) the beatwave force is favored
and rules the injection mechanism (Fig. \ref{fig_f_E_7000_weak_03}).
In this latter case, the beam quality is better without stochastic heating.

\subsection{Very low density: Weak dependance to laser polarization}
\begin{figure}[htbp]
\includegraphics[width=0.25\textwidth]{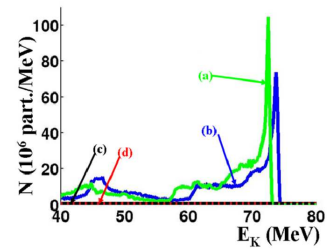}
\caption{Electron energy distribution from 2D PIC simulations.
$a=4$, $a_1=0.1$ and $n_e = 2.5\times 10^{-4} n_c$.
(a): Positive circular polarizations. (b): P linear polarizations.
(c): Negative circular polarizations. (d): S linear polarizations.}
\label{fig_f_E_9000_verylow}
\end{figure}

We now chose parameters relevant for the study of the blowout regime
\cite{malka_2008, martins_2010, lu_prl2006, lu_pop2006}. The plasma density was set to
$n_e = 2.5 \times 10^{-4} n_c$ and the pump pulse intensity to $a=4$, the rest of the simulation
setup was not modified.
Considering two circularly polarized waves rotating in the same direction (positive circular polarizations)
is as efficient as considering P linear polarizations (Fig. \ref{fig_f_E_9000_verylow}).
It means that in this case cold injection \cite{davoine_prl2009, davoine_njp2010}
due to the beating force is the key mechanism governing electron injection.
Stochastic acceleration weakly changes the number and the energy of electrons trapped in the wake field.
The electron momentum distribution along the direction of propagation of the waves during their collision
shows that the effects on electron dynamics of the two polarizations are close, as a result
the electron energy distributions are almost identical (Fig. \ref{fig_f_E_9000_verylow}).
This dominant influence of the beating force upon injection was not clearly stated by
Davoine et \textit{al.} \cite{davoine_prl2009}, here we underline that no relevant difference
is triggered through use of P linear polarizations or positive circular polarizations in the blowout regime.
The next section is devoted to the study of a new means to enhance beam quality after colliding pulse injection,
in the blowout regime.

\section{Influence of a magnetic guide field}	
The influence of a constant homogeneous guide magnetic field on LWFA is studied in this part.
This magnetic field is assumed to be parallel to the direction of propagation of the waves. Electrons are
still externally injected using a colliding counter propagating laser pulse \cite{faure_2006}.
The idea is to guide the electrons in order to improve the quality of the beam which is trapped
in the wake field. This mechanism was first proposed in the context of LWFA, with a single pump laser ($a=3.5$) and an
electronic density $n_e=3\times10^{-3} n_c$ prone to self-injection into the wake field \cite{hur_2008}, with these parameters
self-injection can be dramatically enhanced and beam quality degraded. Here we aim at studying the influence
of a magnetic guide field in the blowout regime \cite{lu_prstab2007} with colliding
pulse injection of the electrons, hence we chose $a = 4$ and
$n_e = 4.4\times10^{17} \textrm{cm}^{-3} = 2.5\times10^{-4} n_c$, thus abiding by $a_0 \ge 4$
and $2\le a_0 \le 2\omega_0/\omega_p$ criteria proposed by Martins et \textit{al.} \cite{martins_2010}.
The required magnetic field necessary to curve electron trajectories is about a hundred teslas \cite{hur_2008}, such values
are particularly strong but still available from the current pulsed magnet technology \cite{lagutin_2003}, the most advanced
magnets can reach 90T for tens of ms durations and centimeter size lengths \cite{zherlit_2010}.
The simulation setup consists in two 30 fs linearly polarized counterpropagating waves with $\lambda=0.8\mu$m wavelength.
They propagate along a constant homogeneous guide field $B_0$ in a cm-long plasma,
the normalised value of $B_0$ is given by $\widetilde{B_0}=e B_0/(m_e \omega_0)$.
Their electric fields are in
the same plane (P linear polarizations). The pump pulse, which creates the accelerating wakefield, is focused
to an $18 \mu$m full width at half maximum. The peak normalized vector potential for this pulse is still $a = 4$.
The low intensity pulse is focused to a $31 \mu$m focal spot at a peak normalized vector potential $a_1 = 0.1$.
The plasma frequency in the presence of a magnetic field can be approximated by
$\omega_m = (\omega_e^2+\Omega^2)^{1/2}$ where $\omega_m$
and $\omega_e$ represent the frequencies of the magnetized and unmagnetized plasma, respectively.
The cyclotron frequency is defined by $\Omega= eB_0/m$. When $B_0 = 125$T and $B_0 = 250$T, one has
$\Omega^2/\omega_e^2 = 0.34$ and $\Omega^2/\omega_e^2 = 1.38$ respectively which proves
that the magnetization of the plasma may have some effect on
the deformation of the wakefield as will be further discussed. Before the collision of the waves the
magnetization has no influence on self-injection, no electron is trapped in the wake.
	
\subsection{Electronic density and transverse currents induced at the rim of the bubble}
\begin{figure}[!htbp]
\begin{center}
  \subfigure[$B=0$]
  {\epsfig{figure=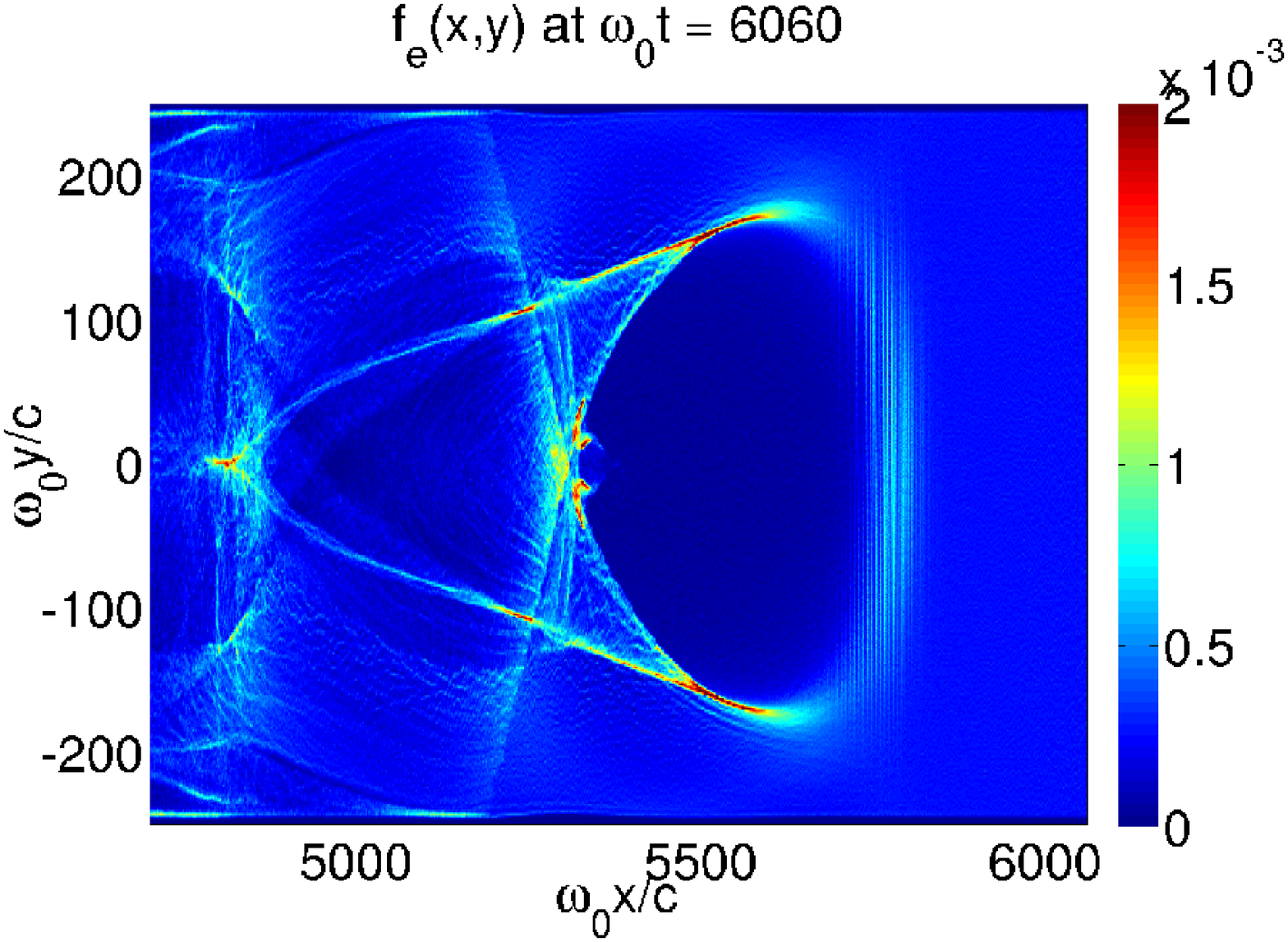,height=2.5cm}
  \label{fig_qxqy_B0_verylow}}
  \subfigure[$B=125$ T]
  {\epsfig{figure=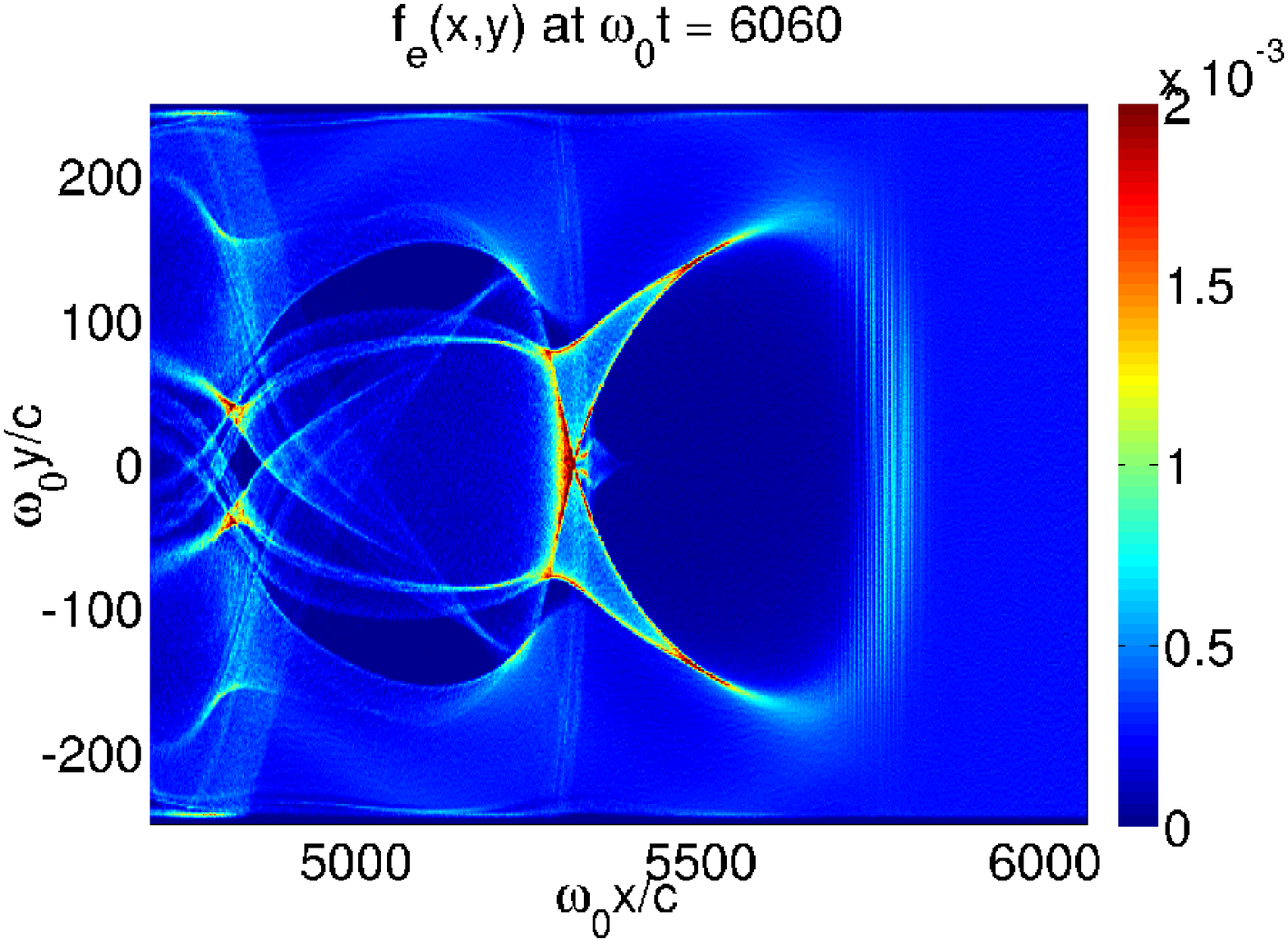,height=2.5cm}
  \label{fig_qxqy_B125_verylow}}
  \subfigure[$B=250$ T]
  {\epsfig{figure=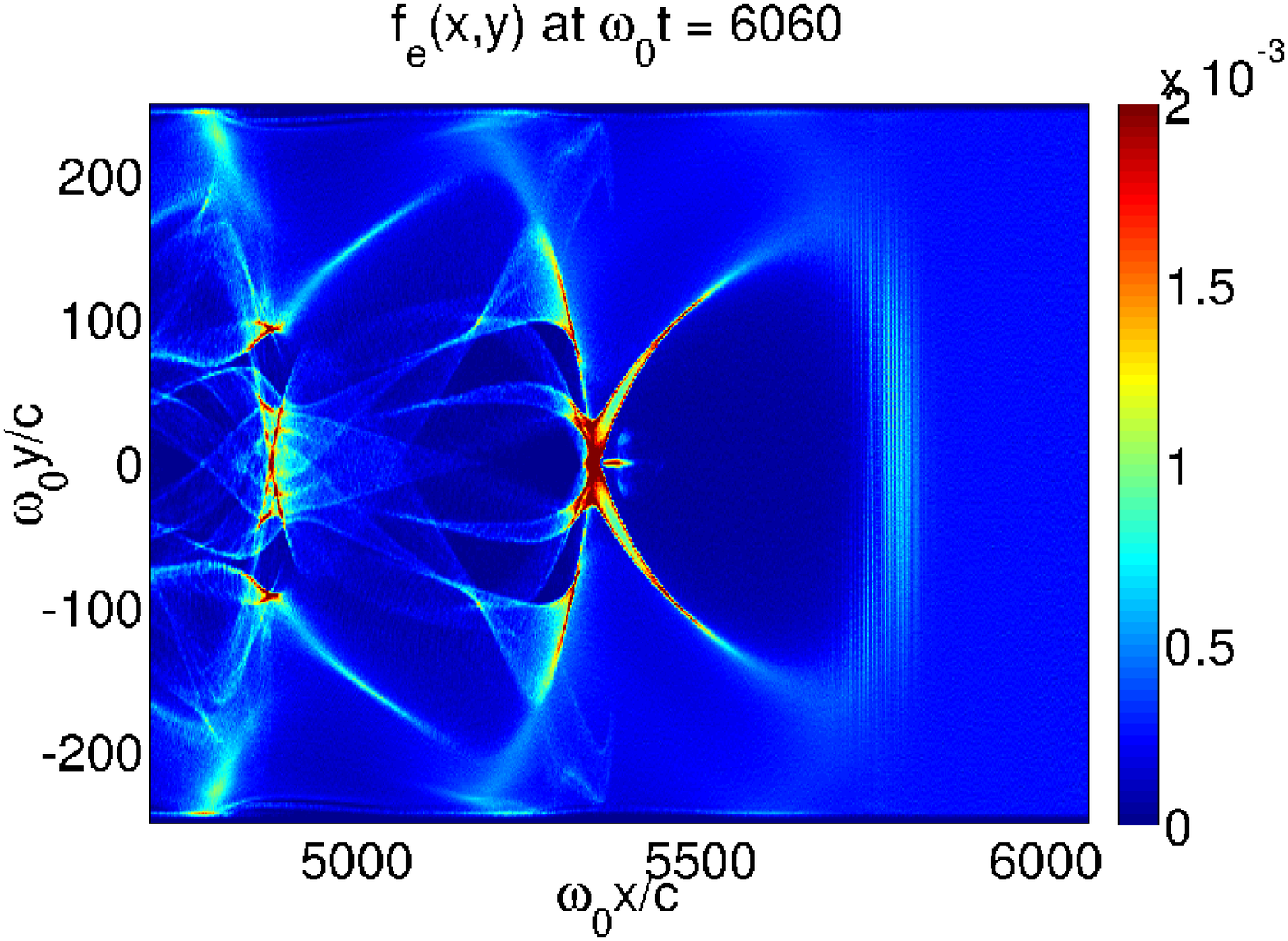,height=2.5cm}
  \label{fig_qxqy_B250_verylow}}
\end{center}
\caption{Electron density long after the collision of the two waves. P linear polarizations.
$a=4$, $a_1=0.1$ and $n_e = 2.5\times 10^{-4} n_c$.
\label{fig_qxqy_verylow}}
\end{figure}

\begin{figure}[!htbp]
\begin{center}
  \subfigure[]
  {\epsfig{figure=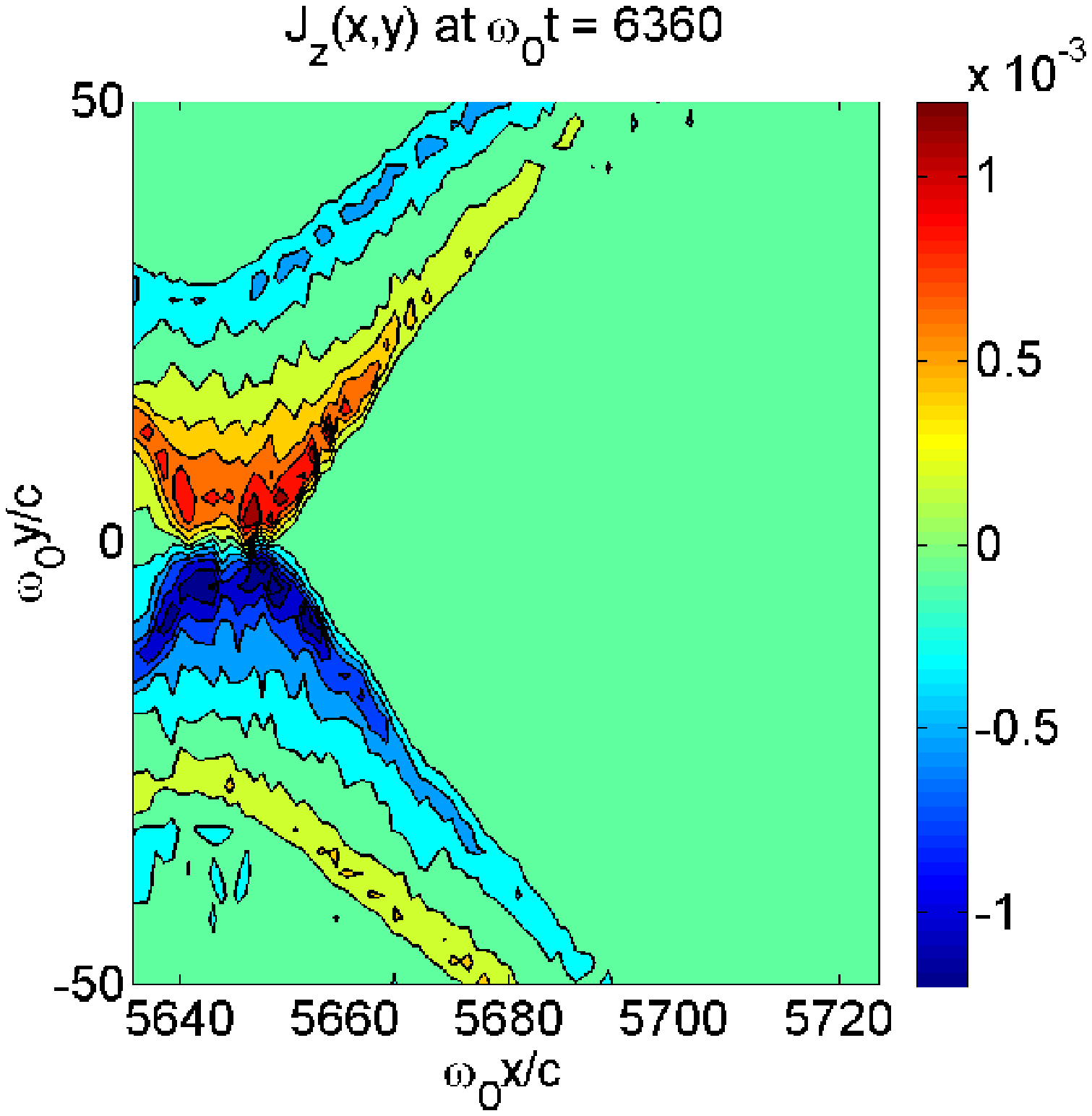,height=3cm}
  \label{fig_cz_6360_verylow}}
%  \subfigure[]
%  {\epsfig{figure=cz_0_b_250_t7260_lin.eps,height=3cm}
%  \label{fig_cz_7260_verylow}}
%  \subfigure[]
%  {\epsfig{figure=cz_0_b_250_t8160_lin.eps,height=3cm}
%  \label{fig_cz_8160_verylow}}
  \subfigure[]
  {\epsfig{figure=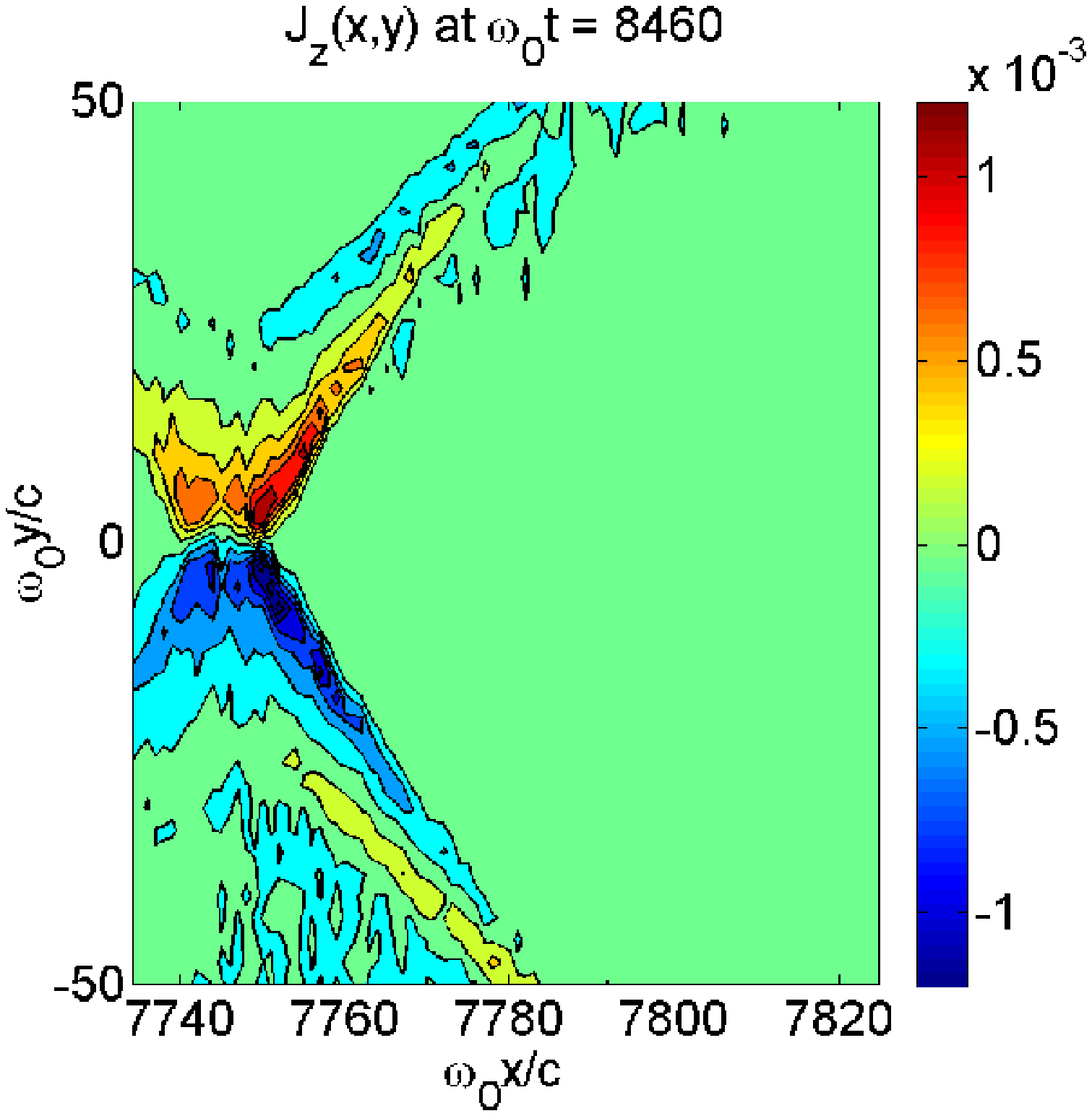,height=3cm}
  \label{fig_cz_8460_verylow}}
\end{center}
\caption{Transverse component of the current density (normalized by $e n_c c$),
the wakefield propagates in a magnetized plasma ($B=250$ T).
$a=4$, $a_1=0.1$ and $n_e = 2.5\times 10^{-4} n_c$.
\label{fig_cz_a4_B250_verylow}}
\end{figure}

Let us first identify the differences brought by the addition of a magnetic guide field to the electronic
distribution at the vicinity of the bubble boundaries (Figs. \ref{fig_qxqy_verylow}). After the electron beam injection
into the wake field, the strong ponderomotive force due to the main pulse still repels electrons and thus
provides them longitudinal and transverse momenta. In the absence of magnetic field, electrons are
submitted to the recall electric field induced by the bubble but the balance between this force and the ponderomotive
force is favorable to the latter ; as a result electrons flee along straight line trajectories
(Fig. \ref{fig_qxqy_B0_verylow}).
When a longitudinal magnetic field is added electrons start to revolve around the bubble as a result of the magnetic force,
this force combined with the electric force induced by the bubble completely modifies the dynamics of the electrons.
The gyro-radius of the electrons with $p_y\ne 0$ is reduced when $B_0$ is raised.
Therefore the corresponding flight path in the $(x,y)$ plane appear to be bent, the trajectory will be even more
curved when the applied field is stronger
(Figs. \ref{fig_qxqy_B125_verylow}-\ref{fig_qxqy_B250_verylow}).
When electrons revolve around the bubble they create a current perpendicularly to the plane of the figure,
as evidenced by Figs. \ref{fig_cz_6360_verylow}-\ref{fig_cz_8460_verylow}.
Given the value of the electronic density at the rear of the bubble ($ 1\times 10^{-3} \le n_e \le 1.5\times 10^{-3}$),
we deduce from Figs. \ref{fig_cz_a4_B250_verylow} the time averaged speed $v_{z}^{\textrm{avg}}$ of electrons revolving around
the bottleneck of the bubble $0.5 \le v_{z}^{\textrm{avg}}/c \le 0.8$.
This current will act as a small solenoid, and thus the longitudinal magnetic field will be
amplified.
This feature and its influence on the beam dynamics will be detailled in the next subsection.

\subsection{A new mechanism to enhance the beam quality}
\begin{figure}[!htbp]
\begin{center}
  \subfigure[]
  {\epsfig{figure=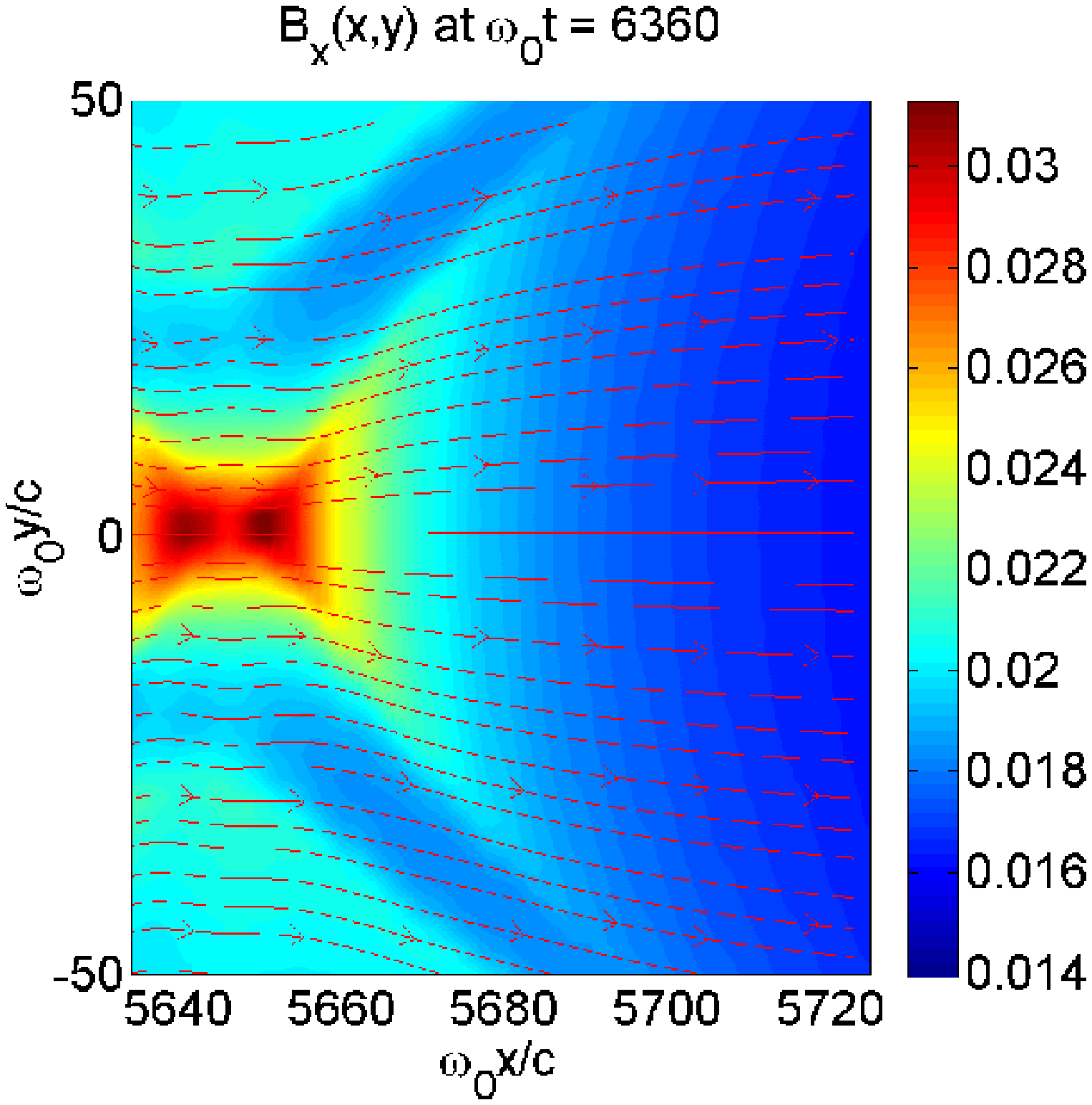,height=3cm}
  \label{fig_bx_6360_verylow}}
%  \subfigure[]
%  {\epsfig{figure=bx_0_b_250_t7260_stream.eps,height=3cm}
%  \label{fig_bx_7260_verylow}}
%  \subfigure[]
%  {\epsfig{figure=bx_0_b_250_t8160_stream.eps,height=3cm}
%  \label{fig_bx_8160_verylow}}
  \subfigure[]
  {\epsfig{figure=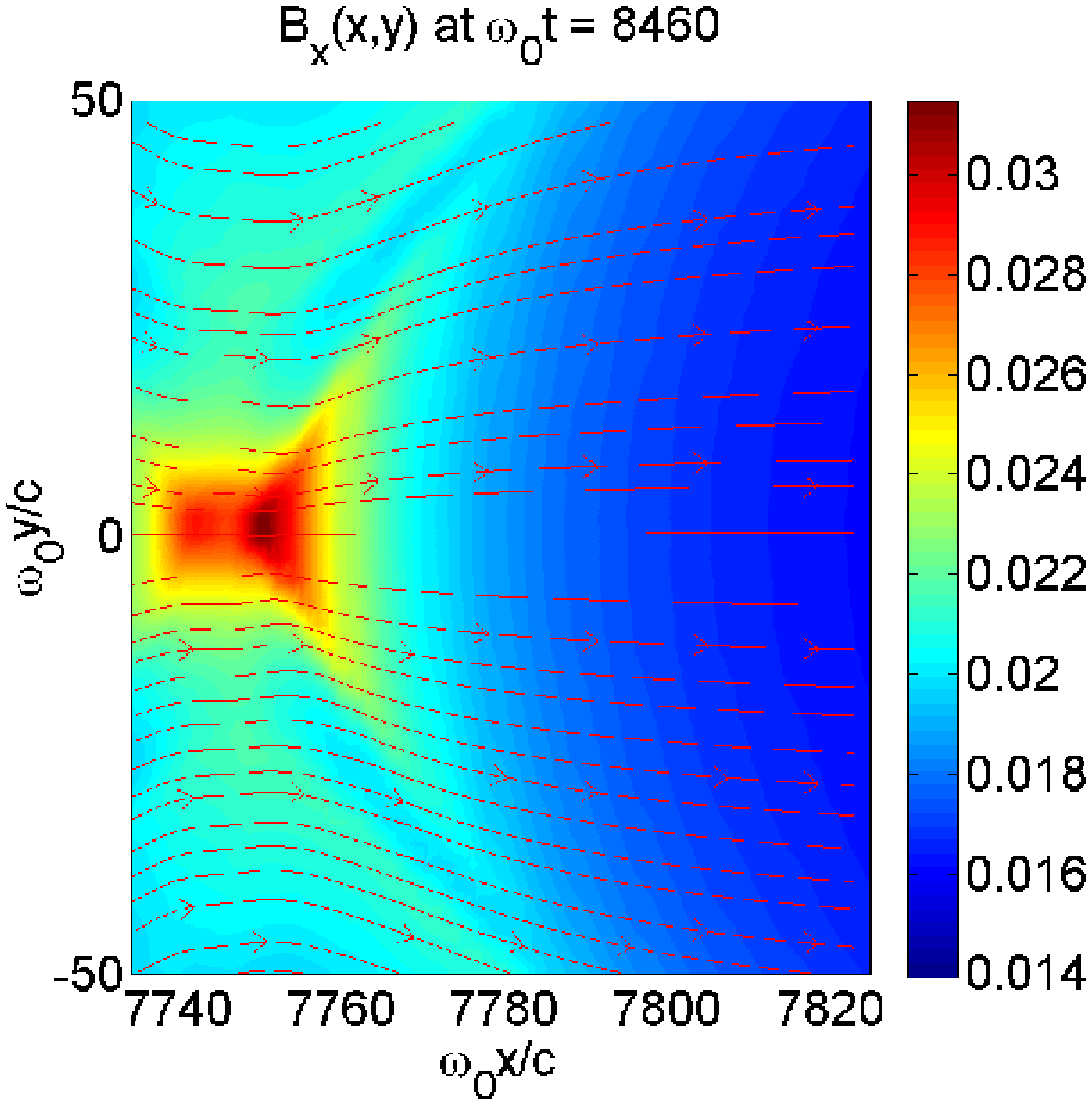,height=3cm}
  \label{fig_bx_8460_verylow}}
\end{center}
\caption{Longitudinal component of the magnetic field (normalized by $m_e \omega_0/e$), the wakefield
propagates in a magnetized plasma $B_0=250$ T (\textit{i.e.} $\widetilde{B_0}\sim 18.7\times 10^{-3}$).
$a=4$, $a_1=0.1$ and $n_e = 2.5\times 10^{-4} n_c$.
Magnetic field lines are superimposed (red curves) on $B_x$ colormap.
\label{fig_bx_a4_B250_verylow}}
\end{figure}

\begin{figure}[!htbp]
\begin{center}
  \subfigure[]
  {\epsfig{figure=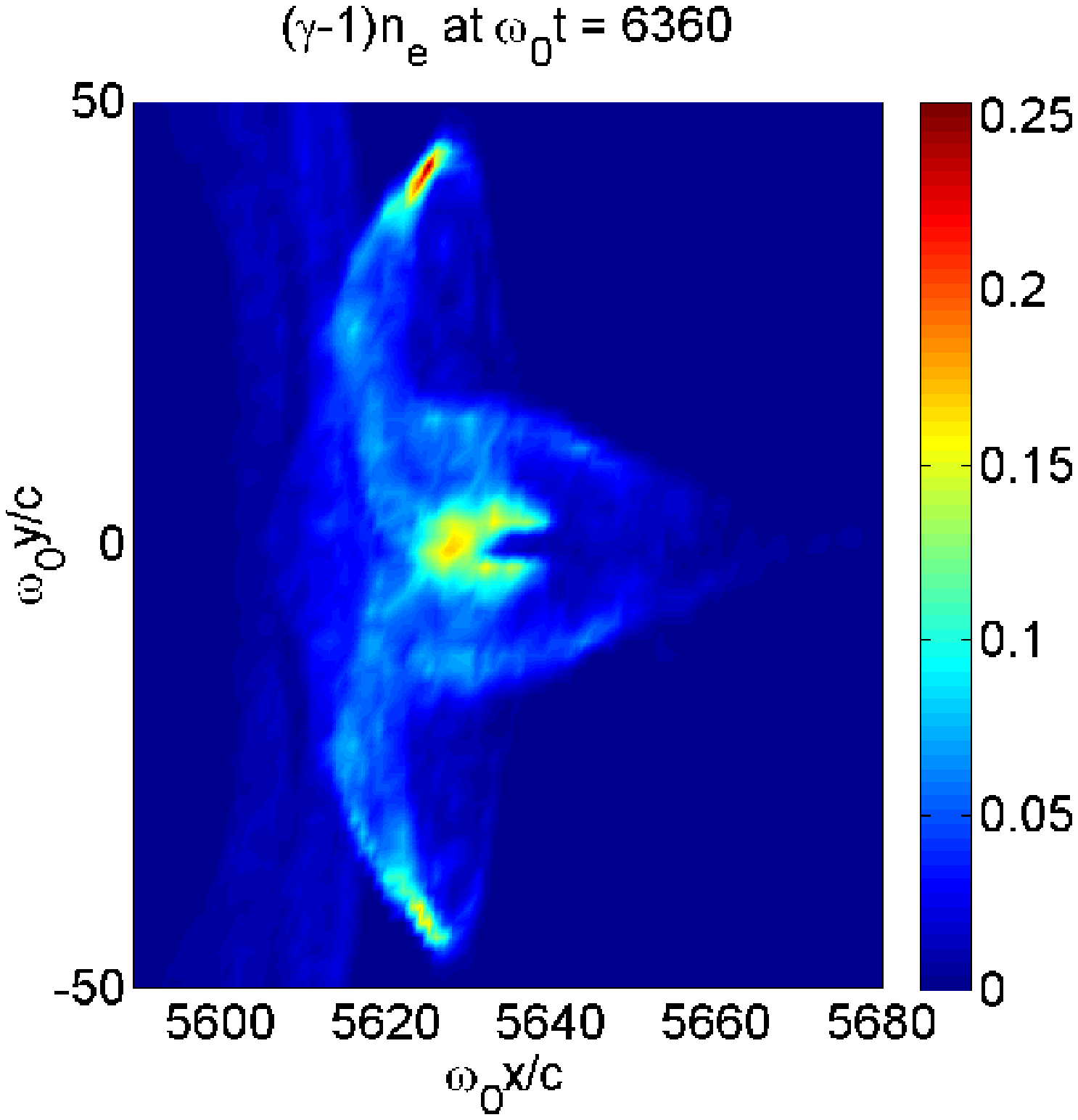,height=3cm}
  \label{fig_gxgy_6360_verylow}}
  \subfigure[]
  {\epsfig{figure=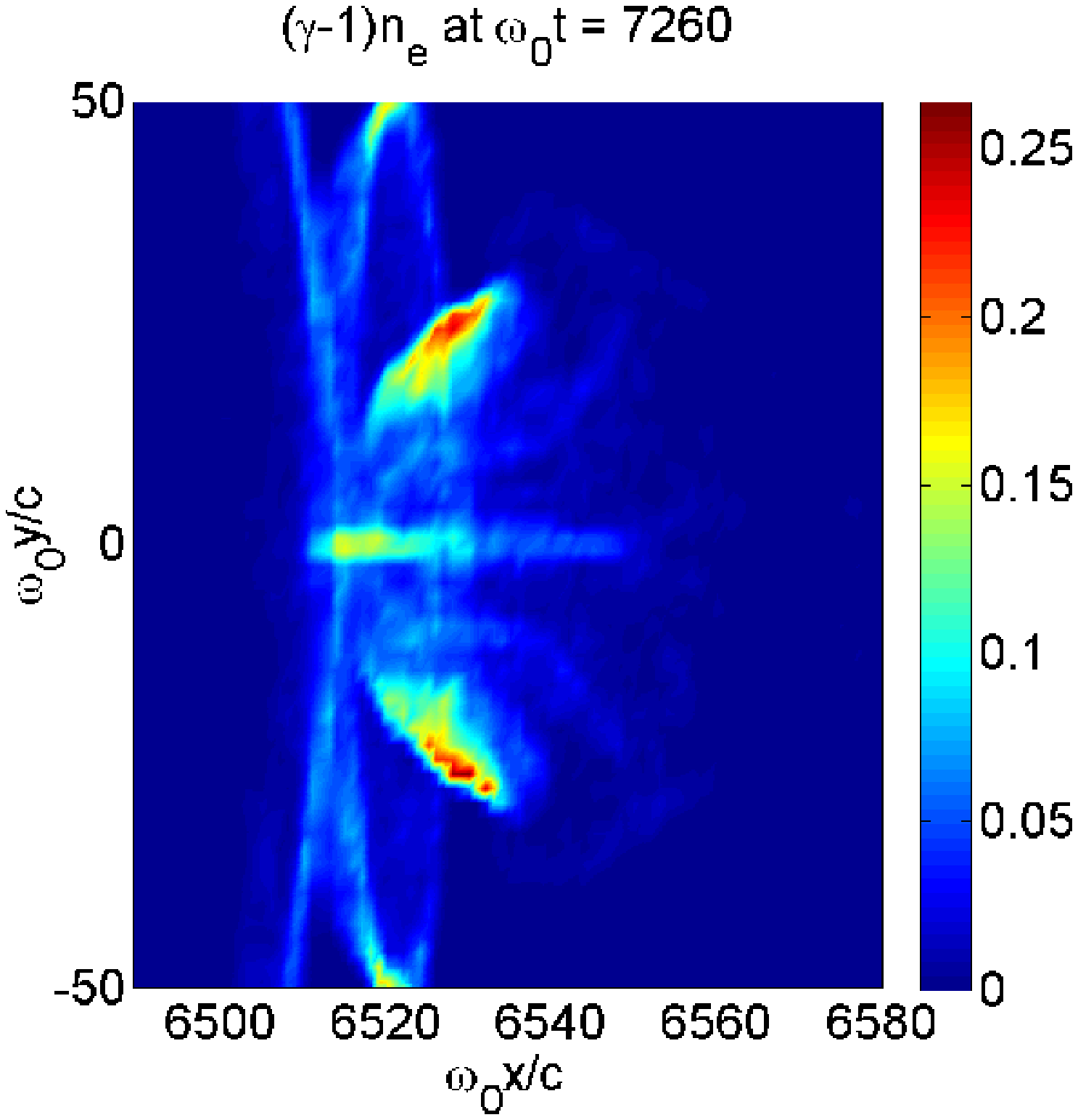,height=3cm}
  \label{fig_gxgy_7260_verylow}}
  \subfigure[]
  {\epsfig{figure=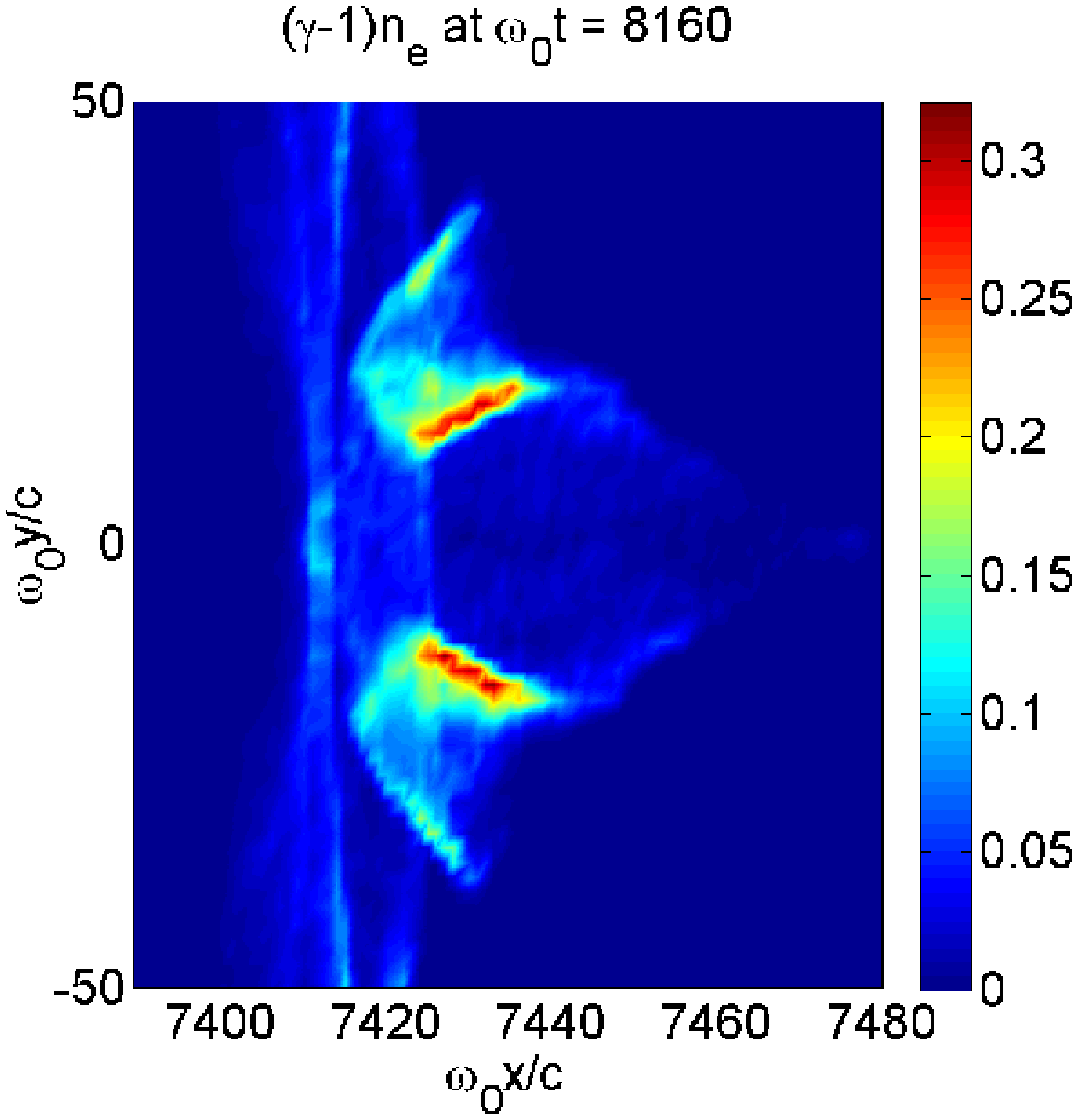,height=3cm}
  \label{fig_gxgy_8160_verylow}}
  \subfigure[]
  {\epsfig{figure=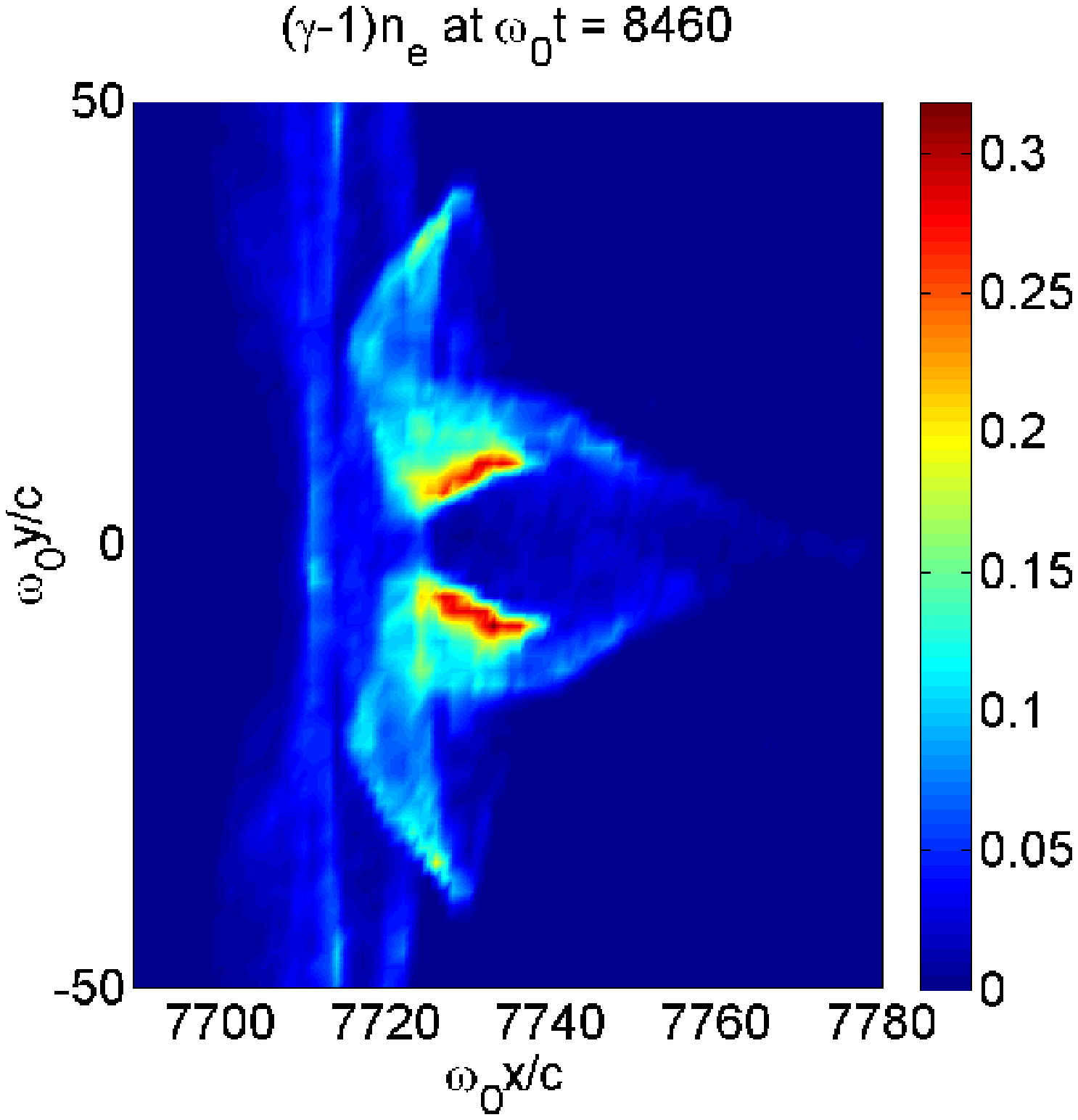,height=3cm}
  \label{fig_gxgy_8460_verylow}}
\end{center}
\caption{Electron kinetic energy density (normalized by $m_e c^2 n_c$), the wakefield
propagates in a non magnetized plasma ($B_0=0$).
$a=4$, $a_1=0.1$ and $n_e = 2.5\times 10^{-4} n_c$.
\label{fig_gxgy_a4_B0_verylow}}
\end{figure}

\begin{figure}[!htbp]
\begin{center}
  \subfigure[]
  {\epsfig{figure=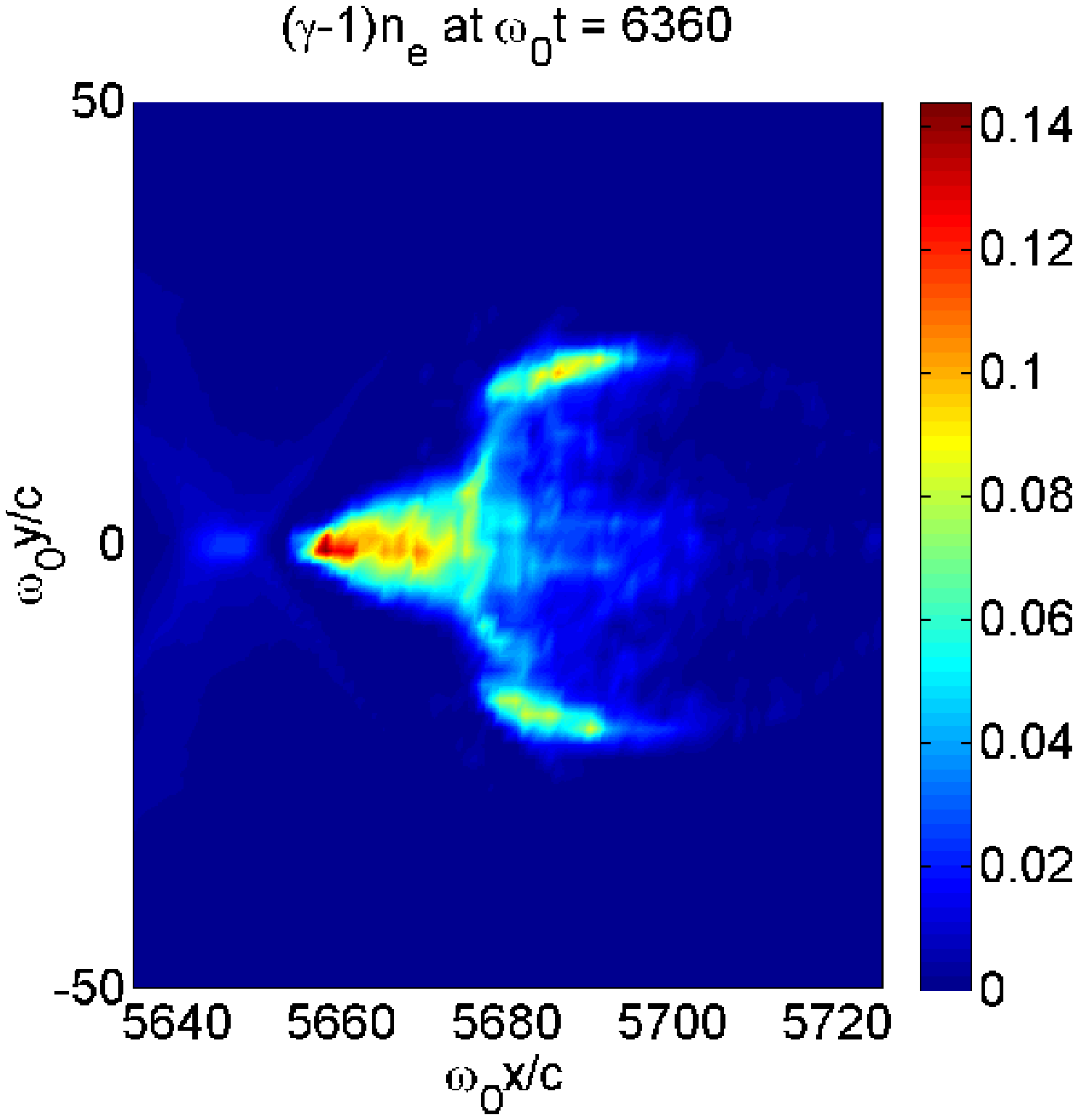,height=3cm}
  \label{fig_gxgy_6360_verylow}}
  \subfigure[]
  {\epsfig{figure=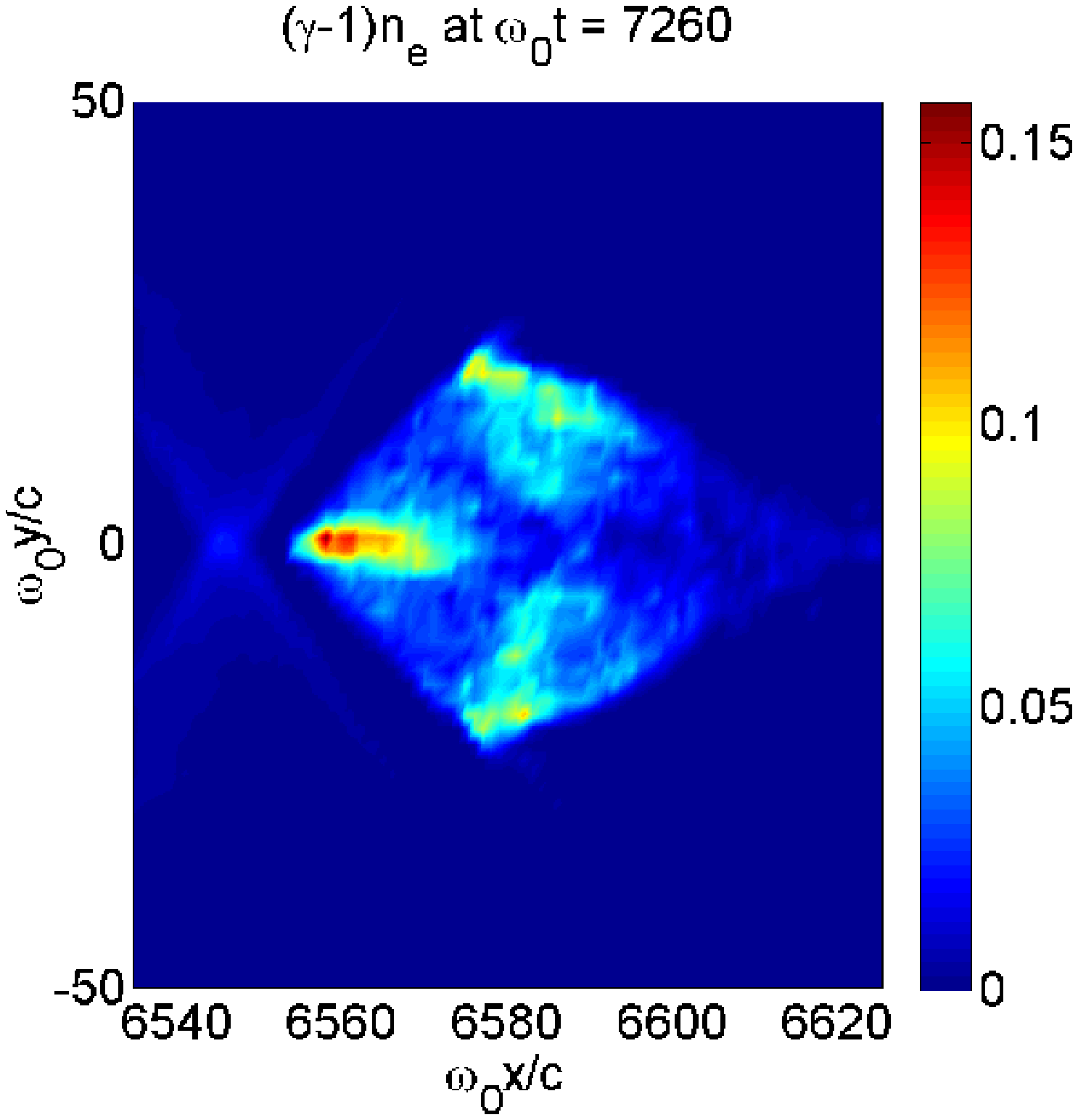,height=3cm}
  \label{fig_gxgy_7260_verylow}}
  \subfigure[]
  {\epsfig{figure=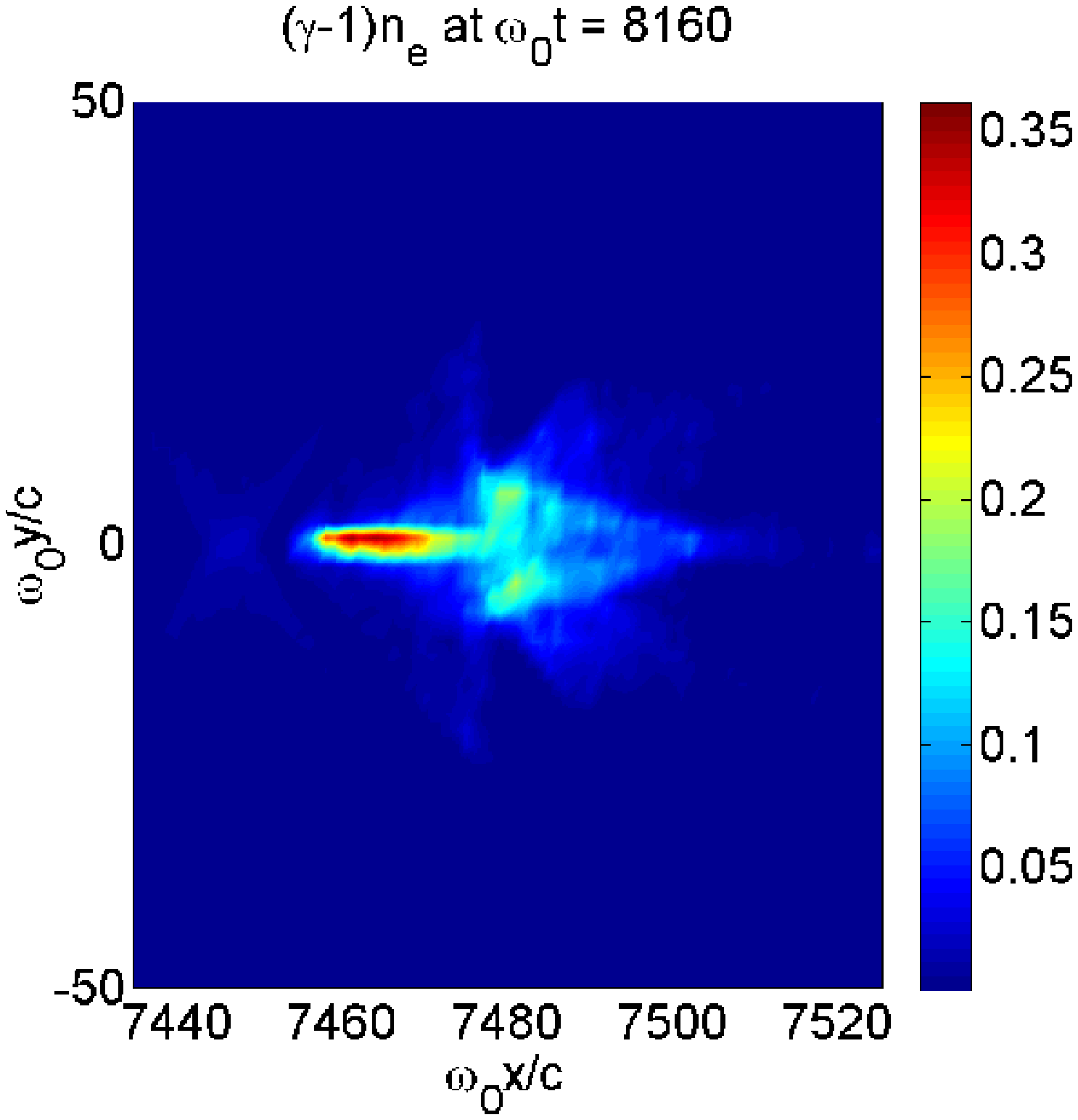,height=3cm}
  \label{fig_gxgy_8160_verylow}}
  \subfigure[]
  {\epsfig{figure=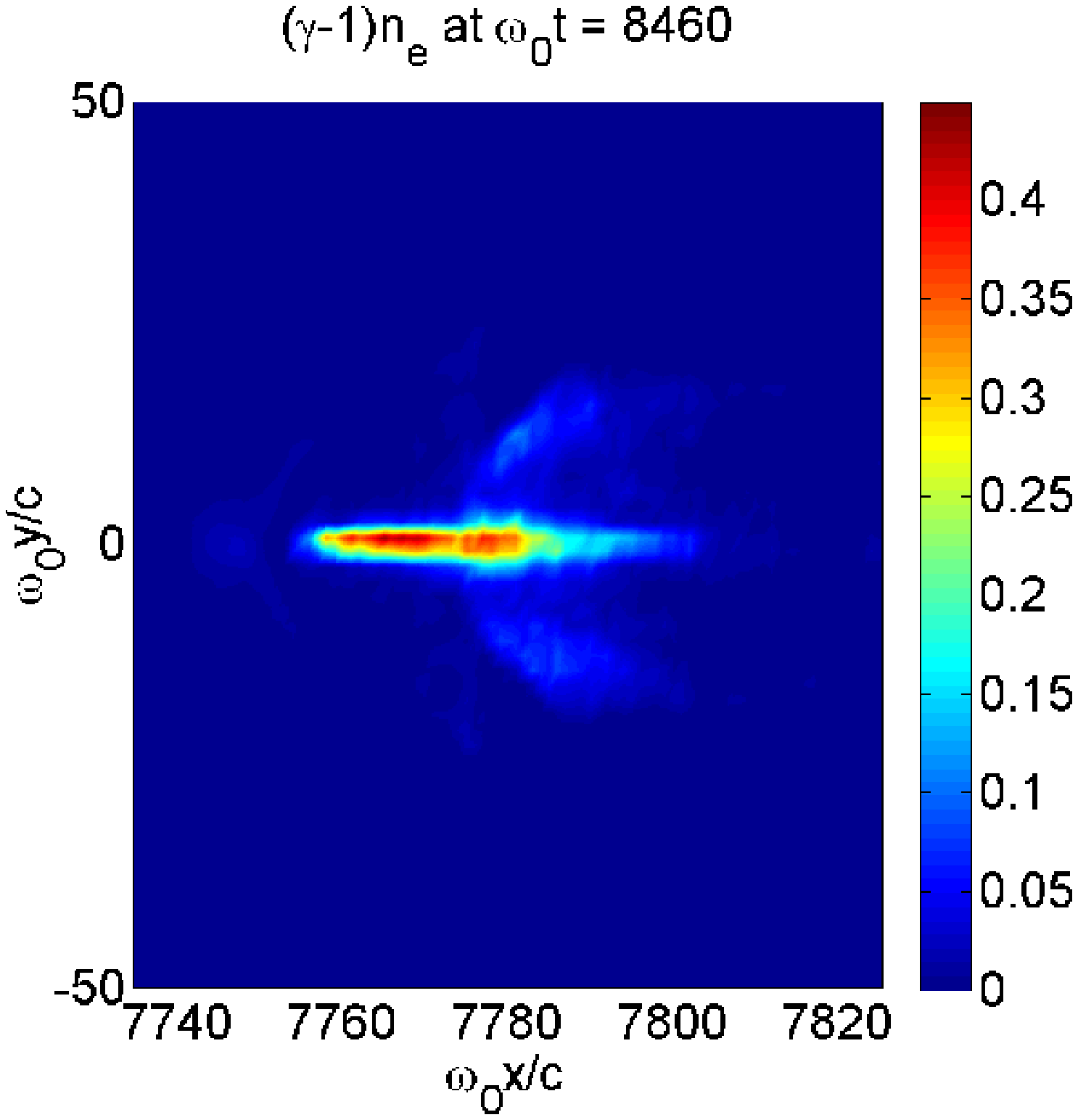,height=3cm}
  \label{fig_gxgy_8460_verylow}}
\end{center}
\caption{Electron kinetic energy density (normalized by $m_e c^2 n_c$), the wakefield
propagates in a magnetized plasma ($B_0=250$ T).
$a=4$, $a_1=0.1$ and $n_e = 2.5\times 10^{-4} n_c$.
\label{fig_gxgy_a4_B250_verylow}}
\end{figure}

The intensity of the magnetic field is almost doubled locally (Figs. \ref{fig_bx_a4_B250_verylow})
compared to the initial ($t=0$) uniform map of $B_x$. We shall underline that this pattern
is stable as we obtain quasi identical maps of $B_x$ in this region of the bubble when the pump
pulse has just entered the plasma around $\omega_0 t =2280$.
Moreover we note that
the geometry of the magnetic field lines is weakly altered by the electronic density
modulations induced by the propagating bubble. Magnetic field lines stay almost
parallel to the propagation direction.
Let us now examine the effect of the magnetic field on the dynamics of the accelerated beam.
To evidence the differences between the magnetized and the unmagnetized regimes, we plotted
kinetic energy density maps showing the evolution of the trapped beam at the rear of the
bubble. In the unmagnetized case (Figs. \ref{fig_gxgy_a4_B0_verylow}), the beam alternatively
explodes (due to space charge effects) and
focalises (due to the focalizing effect of the transverse electric field).
This behavior has a typical $\sim 2000 \omega_0^{-1}$ period. Note that the beam acceleration
is degraded because the components of the splitted beam will not see the maximal value of
the longitudinal electric field (which is located on axis).
The dynamics is completely different in the magnetized case (Figs. \ref{fig_gxgy_a4_B250_verylow}),
the longitudinal magnetic field is strong enough to curve the trajectories and hinder
the explosion of the beam. As a result the beam is almost concentrated on axis, the transverse
emittance is reduced and the main part of the beam, which also corresponds to the region where
the magnetic field is the strongest, always sees the maximum value of the electric field.
Next we will quantify the enhancement of the beam quality through the evolution of
the energy distribution functions of the beam.

\subsection{\label{subsec:beam_quality}Enhancement of the beam quality}

\begin{figure}[!htbp]
\begin{center}
  \subfigure[$B_0=125$ T (bold lines).]
  {\epsfig{figure=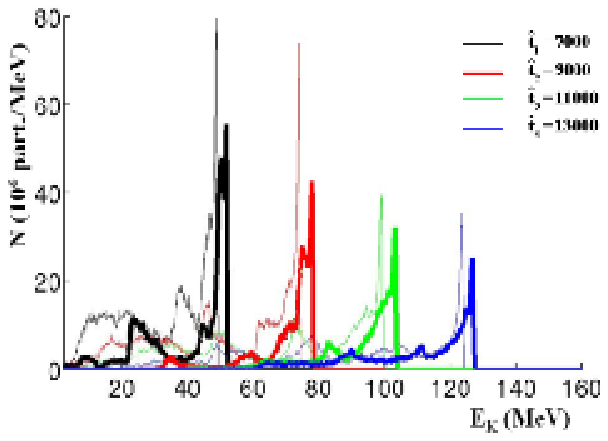,height=2.5cm}
  \label{fig_f_E_B125_verylow}}
  \subfigure[$B_0=250$ T (bold lines).]
  {\epsfig{figure=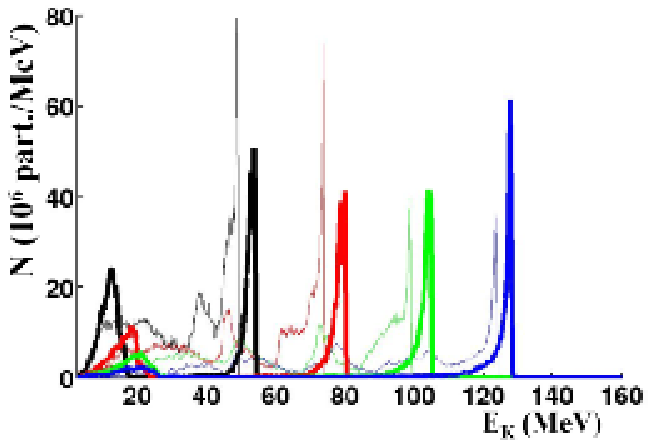,height=2.5cm}
  \label{fig_f_E_B250_verylow}}
\end{center}
\caption{Electron energy distribution from 2D PIC simulations. P linear polarizations.
$a=4$, $a_1=0.1$ and $n_e = 2.5\times 10^{-4} n_c$. Dashed lines correspond to $B=0$.
\label{fig_f_E_a4_verylow}}
\end{figure}

As already mentionned, in the magnetized case, the beam is submitted to a more uniform accelerating field, this pattern boosts the particle acceleration leading to a slightly
higher maximum kinetic energy when compared to the unmagnetized case (Fig. \ref{fig_f_E_a4_verylow}).

\begin{table}
\begin{center}
\begin{tabular}{lcccc}
\hline
\hline
 $ \omega_0 t$ & 7000 & 9000 & 11000 & 13000  \\
 $B_0$ (T) &  &  &    \\
\hline
 0   & 5.33\% & 5.88\% & 4.53\% & 6.61\% \\
     & (2.05\%) & (1.09\%) & (0.81\%) & (0.97\%)  \\
 125 & 5.50\% & 5.45\% & 3.41\% & 3.96\% \\
     & (5.03\%) & (4.89\%) & (1.94\%) & (1.58\%) \\
 250 & 3.96\% & 2.67\% & 2.84\% & 2.91\% \\
     & (3.72\%) & (2.75\%) & (1.54\%) & (0.63\%) \\
\hline
\hline
\end{tabular}
\caption{Evolution of the half width of the electron distribution function with
$a=4$, $a_1=0.1$, $n_e = 2.5 \times 10^{-4}$.
Standard deviation, \textit{i.e.} rms value, and relative variation ($\Delta E_{fwhm}/E_{max}$) of the distribution,
where the subscript $fwhm$ denotes the Full Width at Half Maximum.}\label{tab_dEsE_a4}
\end{center}
\end{table}

The focalizing magnetic field reduces the low energy tail of the energy spectrum, as can be seen in Figs \ref{fig_f_E_a4_verylow}. Obviously, this trend is enhanced when the guide field rises.
With no guide field, the relative variation of the energy at full width at half maximum (fwhm)
$\Delta E_{fwhm} /E_{max} \approx 1\%$ is excellent, but the rms value of the energy spread has small variations and reaches ~7\% at
the end of the simulation (Table \ref{tab_dEsE_a4}). When $B_0 = 125$ T, on the one hand the spread of the low energy tail of the
distribution is reduced as shown by Fig. \ref{fig_f_E_B125_verylow}, and confirmed by the rms value $\sim 4\%$, but on the other hand
$\Delta E_{fwhm} /E_{max}$ is slightly degraded. When $B_0 = 250$ T, untrapped electrons carrying energies about 10 Mev concentrate
($n_e$ locally reaches $1.5\times 10^{-3} n_c$) at the rear of the bubble. These low energy (i.e. $0 < E_K < 25$ Mev) electrons
are evidenced by bumps in the beam energy distribution (Fig \ref{fig_f_E_B250_verylow}). This low energy bump slowly slides out of the
simulation box as these electrons are not injected in the wakefield, and therefore should not be considered for
the interpretation of the diagnostics concerning the accelerated beam. According to this comment, we note that a
250 T guide field is enough to completely suppress the low energy tail during the whole simulation.
A clear enhancement of the beam quality is obtained, first the final rms value is below 3\% and $\Delta E_{fwhm} /E_{max} < 1\%$
(table \ref{tab_dEsE_a4}) thus providing a very sharp control on the final energy of the beam, and second the number of electrons
at the highest energies does not vanish, as in the unmagnetized case, but on the contrary grows up to $60\times 10^6$ part/Mev,
nearly twice the value of the unmagnetized case ! To our knowledge such an acute
mono-energetic electron beam production, with complete extinction of the low energy tail has never been evidenced.
	
%\begin{figure}[!htbp]
%\begin{center}
%  \subfigure[$B=0$]
%  {\epsfig{figure=qxpx_7020_2e-4_B0.eps,height=2.5cm}
%  \label{fig_qxpx_B0_verylow}}
%  \subfigure[$B=125$ T]
%  {\epsfig{figure=qxpx_7020_2e-4_B125.eps,height=2.5cm}
%  \label{fig_qxpx_B125_verylow}}
%  \subfigure[$B=250$ T]
%  {\epsfig{figure=qxpx_7020_2e-4_B250.eps,height=2.5cm}
%  \label{fig_qxpx_B250_verylow}}
%\end{center}
%\caption{Electron longitudinal momemtum distribution from 2D PIC simulations long after the %collision of the
%two pulses. P linear polarizations. $a=4$, $a_1=0.1$ and $n_e = 2.5\times 10^{-4} n_c$.
%\label{fig_qxpx_a4_verylow}}
%\end{figure}
	
For a given pump pulse, two well-known controllers of the low intensity laser can be used to optimize the
beam quality. The low intensity pulse duration can be monitored and the transverse fwhm of the spatial
envelope adjusted, these two parameters together usually make it possible to get a quasi-mono-energetic electron
beam in the blow-out regime \cite{martins_2010, davoine_prl2009}.
There are other ways to enhance the beam quality. For example, one can resort to a longitudinal gradient of the electronic density to enhance trapping
\cite{davoine_prl2009, geddes_2004, faure_pop2010}.
In an alternate approach, assuming initially homogenous plasma, one can slowly evolve the laser pulse shape
to alternate periods of expansion and contraction of the bubble, to respectively trigger and stop
self-injection \cite{kalmykov_2011} of the electrons into the bubble. However this technique seems hard to adjust
to get a unique mono-energetic bunch. In this paper, injected electrons are confined by using a magnetic guide field but
we note that the intensity of the field, required to substantially enhance the beam quality, depends on lasers pulse shapes and durations. In this section we did not pay attention
to the tuning of the low intensity colliding pulse, we have shown that the guiding induced by $B_0$ is enough
beyond some threshold, function of the parameters of the simulation. However, if we decide to lower the blowout
stability by increasing the main pulse intensity, the required intensity of the guide field grows to values
largely out of reach of the current technology, then fine-tuning of the colliding pulse becomes necessary.
The next section is devoted to this issue.

\section{Influence of a magnetic guide field at higher intensities}

\begin{figure}[!htbp]
\begin{center}
\subfigure[$\omega_0 t = 13000$]
{\epsfig{figure=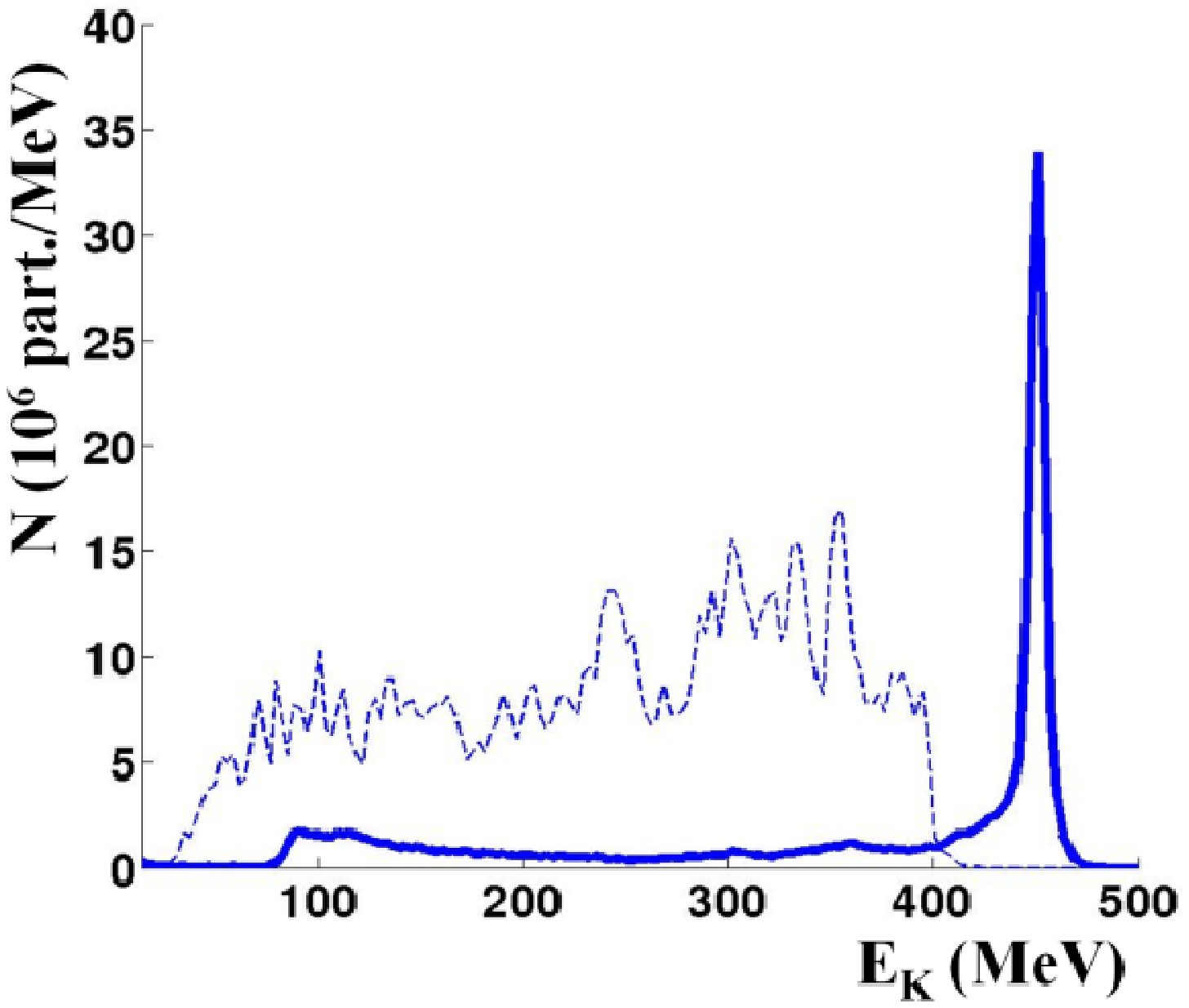,height=3cm}
\label{fig_f_E_a10_13000}}
\subfigure[$\omega_0 t = 30000$]
{\epsfig{figure=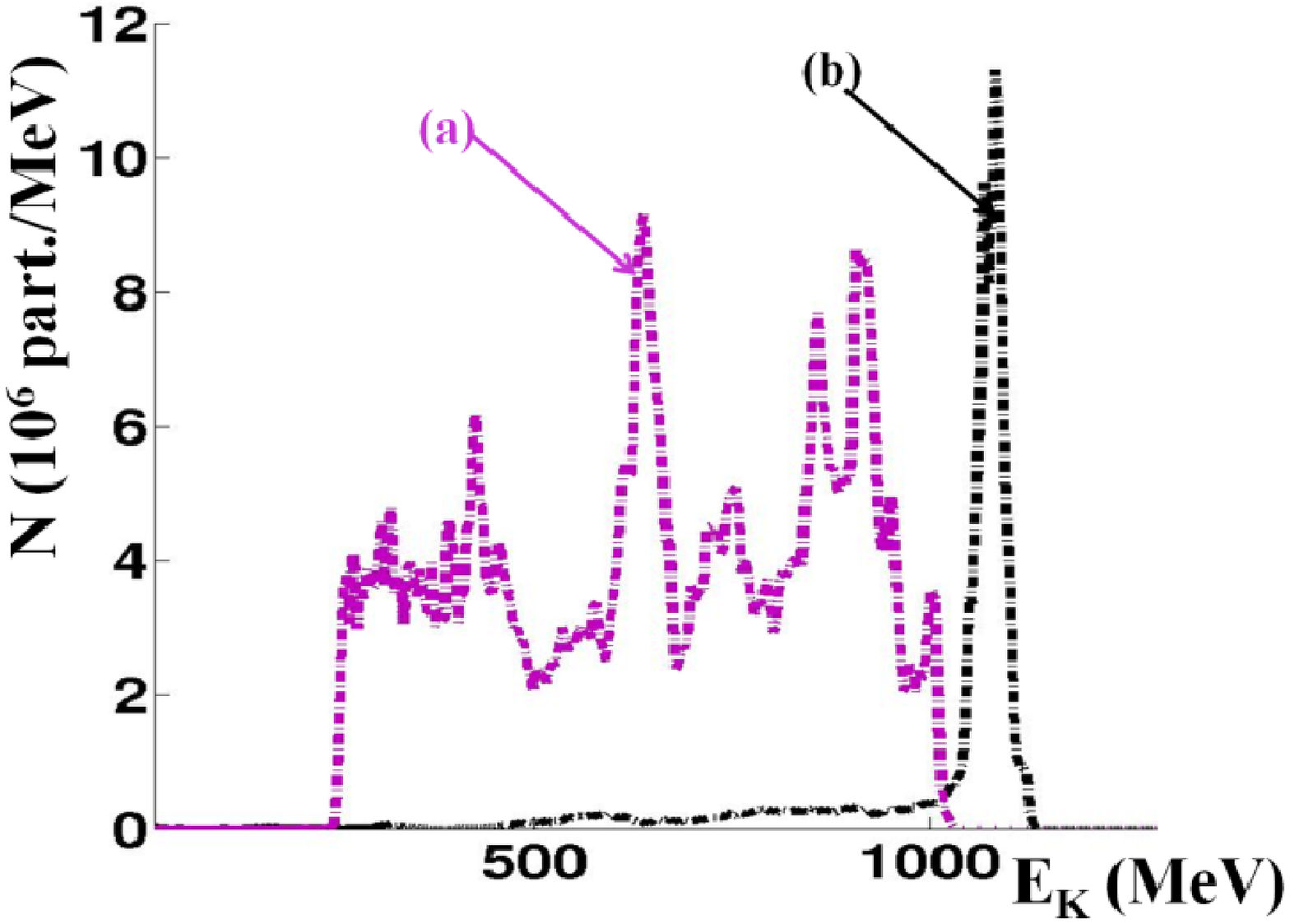,height=3cm}
\label{fig_f_E_a10_30000}}
\end{center}
\caption{Electron energy distribution from 2D PIC simulations with
$a=10$, $a_1=0.1$, $n_e = 2.5\times 10^{-4} n_c$ and $D=10 \mu$m.
The dashed line corresponds to $B_0=0$ and the bold one to $B_0=125$ T.
\label{fig_f_E_a10_verylow}}
\end{figure}

A very high intensity wave is considered now, the pump pulse which is linearly polarized is assumed to have a
peak normalized intensity $a = 10$, and a duration of 30 fs. It is focused to a $35.65 \mu$m large focal spot.
The colliding pulse has a peak normalized intensity $a_1 = 0.1$, a 30fs duration and three focal spot sizes were
considered $D = 60 \mu $m, 36 $\mu$m and 10 $\mu$m. The wavelength of the waves is $\lambda= 0.8 \mu$m. In these simulations, the
plasma density is still $n_e = 2.5\times10^{-4} n_c$. Unless otherwise mentionned the following simulations were run with $B_0=125$T.
When considering $a = 10$ and a counter-propagating wave focused to a 10 $\mu$m focal spot, one has a paramount
effect of the magnetic field on the distribution function (Table \ref{tab_dEsE_a10}, Fig. \ref{fig_f_E_a10_verylow}). The electron energy distribution
becomes almost mono-energetic. After the laser pulse has propagated through the plasma by 3.8 mm, the electron energy distribution is still
quite mono-kinetic and the maximum electronic energy exceeds one GeV (Fig. \ref{fig_f_E_a10_30000}).
Table \ref{tab_dEsE_a10} shows the evolution of the energy spread of the electron energy distribution with time.
When no field is applied the quality decreases whereas we get an acute control on the beam energy with $B_0$.
It must be pointed out that the accelerated charge is small (close to 50 picocoulombs).
The electron energy distributions corresponding to four focal spots are compared at some time.
Figure \ref{fig_f_E_a10_D10_dt10fs_B250} shows that the distribution becomes much more mono-energetic when the focal spot of the
perturbing wave is smaller than the one of the main pulse. The magnetic field is much more efficient
when considering a small value of D.
Then, we have checked that this peak exists in a very small range of values close to $a_1 = 0.1$.
When $a_1 = 0.08$ the distribution function shows a lower magnitude peak, and the charge accelerated in the
first bubble is about 10 picocoulombs. The peak still exists when $a_1 = 0.102$.
But in the case of a higher value of $a_1$(for instance $a_1 = 0.15$) the energy distribution does not show a very thin high-energy peak any longer
because the injected charge is higher and the magnetic field is too weak to concentrate the beam efficiently.

\begin{figure}[htbp]
\includegraphics[width=0.25\textwidth]{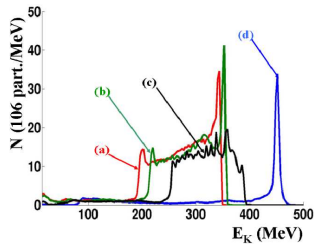}
\caption{Electron energy distribution from 2D PIC simulations at $\omega_0 t=13000$ with
$a=10$, $a_1=0.1$, $n_e = 2.5\times 10^{-4} n_c$ and $B_0=125$ T.
(a)$D=59.5\mu$m, (b)$D=35.7\mu$m, (c)$D=20\mu$m, (d)$D=10\mu$m.
\label{fig_f_E_a10_D10_dt10fs_B250}}
\end{figure}

\begin{figure}[htbp]
\includegraphics[width=0.25\textwidth]{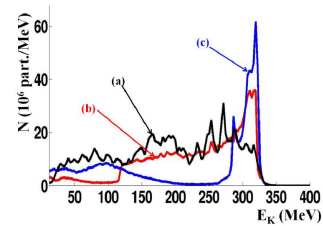}
\caption{Electron energy distribution from 2D PIC simulations at $\omega_0 t=13000$  with
$a=10$, $a_1=1$, $n_e = 2.5\times 10^{-4} n_c$, $D=10 \mu$m and $\Delta t=30$fs.
(a)$B_0=0$, (b)$B_0=125$ T, (c)$B_0=250 T$.
\label{fig_f_E_a10_B0-125-250}}
\end{figure}

\begin{figure}[htbp]
\includegraphics[width=0.25\textwidth]{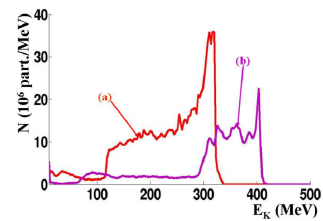}
\caption{Electron energy distribution from 2D PIC simulations at $\omega_0 t=13000$ with
$a=10$, $a_1=1$, $n_e = 2.5\times 10^{-4} n_c$, $B_0=125$ T and $D=10 \mu$m.
(a)$\Delta t=30$fs, (b)$\Delta t=10$fs.
\label{fig_f_E_a10_dt10-30fs_B125}}
\end{figure}

\begin{table}
\begin{center}
\begin{tabular}{lcccc}
\hline
\hline
 $ \omega_0 t$ & 7000 & 13000 & 22000 & 30000  \\
 $B_0$ (T) &  &  &    \\
\hline
 0   & 40.5\% & 41.2\% & 36.0\% & 33.2\% \\
     & (23.0\%) & (86.1\%) & (59.3\%) & (100.9\%)  \\
 125 & 6.4\% & 1.7\% & 5.0\% & 3.35\% \\
     & (4.85\%) & (1.3\%) & (2.0\%) & (2.6\%) \\
\hline
\hline
\end{tabular}
\caption{Evolution of the half width of the electron distribution function.
$a=10$, $a_1=0.1$, $n_e = 2.5 \times 10^{-4}$.
Standard deviation,\textit{i.e.} rms value, and relative variation ($\Delta E_{fwhm}/E_{max}$)
of the distribution.}\label{tab_dEsE_a10}
\end{center}
\end{table}
No high-energy peak was seen in the distribution function for larger values of $a_1$, actually the same kind of
distribution is obtained when $a_1 = 0.2$, $a_1 = 0.5$ and $a_1 = 1$.
Figure \ref{fig_f_E_a10_B0-125-250} shows that the electron energy distribution becomes more mono-energetic when the magnitude of the
magnetic guide field is increased to 250 T, accordingly with the results obtained with a lower intensity of the pump pulse
($a_0 = 4$) in section \ref{subsec:beam_quality}.
Figure \ref{fig_f_E_a10_dt10-30fs_B125} shows that, a shorter pulse duration ($\Delta t = 10$ fs) for the counterpropagating wave, also makes the
electron energy distribution more mono-energetic.
Then, in order to obtain a more mono-energetic distribution when $a = 10$ and $a_1 = 1$, a very strong magnetic
field $B_0 =250 T$ and a short duration for the counterpropagating wave $\Delta t = 10$ fs were considered (Fig. \ref{fig_f_E_a10_D10_dt10fs_B250}).
As expected, a high energy peak is obtained.
One should point out that the charge accelerated in the first bubble is still close to 50 picocoulombs.
To summarize when $a_1$ is too high, that is when the charge injected in the wakefield exceeds some treshold,
we can not stop a drop of the beam quality by imposing an external field alone, at least the main parameters
(focal spot size and duration) of the colliding pulse shall be reduced.

\begin{figure}[htbp]
\includegraphics[width=0.25\textwidth]{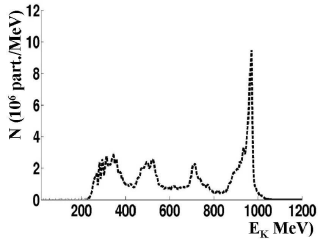}
\caption{Electron energy distribution from 2D PIC simulations at $\omega_0 t=30000$ with
$a=10$, $a_1=1$, $n_e = 2.5\times 10^{-4} n_c$ and $B_0=250$ T.
The spatio-temporal shape of the laser enveloppe is defined by $D=10 \mu$m and $\Delta t=10$fs.
\label{fig_f_E_a10_D10_dt10fs_B250}}
\end{figure}

\section{Conclusions}
In the first part of this paper, we summarized some essential results about the most efficient choice of
polarizations for injection in the bubble, in the colliding pulse scheme.
At rather high electron density ($\gtrsim 10^{-3} n_c$) and moderately relativistic electromagnetic wave
intensity ($a_0\lesssim 2$), more
particles are accelerated to high energies in the case of P linear polarizations that is to say when electrons
undergo the action of the beatwave force and all the others. For higher intensities and lower densities
($\sim 10^{-4} n_c$) the beatwave force, that is cold injection, can be more efficient.
The second and main part of this paper has been devoted to the study of the influence of an external
static magnetic field on the wakefield acceleration process, within the colliding pulse scheme.
To our knowledge this idea was never explored.
The magnetic field is supposed parallel to the direction of propagation of the two counter-propagating waves.
It has been shown that the $B_0$ field creates a transverse current, the latter current can
induce a raise of $B_x$ at the rear bottleneck of the bubble.
Therefore the beam dynamics is substantially modified as the beam is constrained to stay in the maximum
acceleration region of the bubble.
Beam emittance is considerably reduced and maximum kinetic energy slightly boosted compared to the
unmagnetized case. This mechanism provides means to dramatically enhance the beam quality in the blowout
regime. We achieved tremendous amelioration with the setup: $a=10$, $a_1=0.1$ and $n_e=2.5 \times 10^{-4} n_c$.
After roughly 4 mm of wakefield acceleration, without $B_0$ field the electronic energy distribution is noisy
$\Delta E_{fwhm}/E_{max}\sim 100\%$ whereas we get $\Delta E_{fwhm}/E_{max}\lesssim 3\%$ when the plasma
is magnetized with a 125 T field.
Nevertheless the intensity of the $B_0$ field may be limited by technological considerations \cite{zherlit_2010},
thus acute control of the beam quality may require some fine-tuning of the colliding pulse parameters.
For a given pump pulse, one should adapt the intensity, duration and focal spot size of the
counter-propagating laser pulse.

%\appendix
%\section{}

%\begin{acknowledgments}
%\end{acknowledgments}

\bibliography{prz_drouin}

\end{document}